\documentclass[aps,prc,floatfix,showpacs,twocolumn]{revtex4-1}
\usepackage{ulem}
\usepackage{dcolumn}
\usepackage{bm}
\usepackage{color}
\usepackage{amssymb}
\usepackage{amsmath}
\usepackage{graphicx}
\usepackage{amsfonts}
\usepackage{slashed}
\usepackage{float}
\usepackage{hyperref}
\usepackage{array}
\allowdisplaybreaks
\usepackage{dcolumn}
\usepackage{epsf}
\usepackage{slashed}
\begin{document}

%\title{Pion-production in lepton-nucleus scattering within the extended factorization scheme}
\title{{\bf Electroweak pion-production on nuclei within the extended factorization scheme}}
\author{
{Noemi} Rocco$^{\, {\rm a,b} }$,
{Satoshi X.} Nakamura$^{\, {\rm c, d} }$,
{T.-S. H.} Lee$^{\, {\rm a} }$,
{Alessandro} Lovato$^{\, {\rm a, e} }$
}
\affiliation{
$^{\,{\rm a}}$\mbox{Physics Division, Argonne National Laboratory, Argonne, Illinois 60439, USA}\\
$^{\,{\rm b}}$\mbox{Theoretical Physics Department, Fermi National Accelerator Laboratory, P.O. Box 500, Batavia, IL 60510, USA}\\
$^{\,{\rm c}}$\mbox{University of Science and Technology of China, Hefei 230026, 
People's Republic of China}\\
$^{\,{\rm d}}$\mbox{State Key Laboratory of Particle Detection and
Electronics (IHEP-USTC), Hefei 230036, People's Republic of China}\\
$^{\,{\rm e}}$\mbox{INFN-TIFPA Trento Institute of Fundamental Physics and Applications, Via Sommarive, 14, 38123 Trento, Italy}\\
}
\date{\today}

\begin{abstract} 
We have applied the extended factorization scheme to investigate the electroweak pion production on nuclei. The ANL-Osaka model, which was obtained by analyzing the data of $\pi N$, $\gamma N$, $N(e,e'\pi)$ and $N(\nu,\mu\,\pi) $ reactions up to invariant mass $W=$ 2 GeV, is used to generate the matrix elements of current operators relevant to pion-production off the nucleon. Medium effects on the $\Delta$ (1232) component of meson-exchange current are included by using a $\Delta$-nucleus potential determined from the previous $\Delta$-hole model studies of pion-nucleus reactions. Nuclear correlations in the initial target state and in the spectator system(s) are modeled using realistic hole spectral functions. As a first step, we show that the data of $^{12}$C$(e,e')$ up to the $\Delta$ (1232) region can be described reasonably well. The interplay between the pion production and two-body meson-exchange mechanisms is shown to be essential in improving the agreement with the data in the ``dip'' region, between the quasielastic and the $\Delta$ (1232) peaks.
Predictions for $^{12}$C$(\nu,\mu\,\pi)$ have also been made. They can be used to estimate pion-emission rates in neutrino-nucleus cross section, which constitutes an important systematic uncertainty to the reconstructed neutrino energy. With further improvements of the Metropolis Monte-Carlo techniques to account for final states comprised of more than two particles, our approach can be employed up to $W=$ 2 GeV, where two-pion production and higher mass nucleon resonances must be included for analyzing the data from accelerator-based neutrino-oscillation experiments.
\end{abstract}
\maketitle
%%%%%%%%%%%%%%%%%%%%%%%%%%%%%%%%%%%%%%%%%%%%%%%%%%%%%%%%%%%%%%%%%%%%%%%%%
\section{Introduction}
The development of the world-wide accelerator-based neutrino-oscillation program has been a springboard for advancing the theoretical description of lepton interactions with nuclei~\cite{Benhar:2015wva,Katori:2016yel,Alvarez-Ruso:2017oui}. Oversimplified models of nuclear dynamics, such as the relativistic Fermi Gas, have proven to be inadequate to reproduce quasi-elastic charge-changing scattering data on $^{12}$C~\cite{AguilarArevalo:2007ab,AguilarArevalo:2010zc}. As a result, more sophisticated approaches, capable of providing a rather accurate description of available inclusive neutrino scattering data have been devised~\cite{Leitner:2008ue,Benhar:2010nx,Martini:2010ex,Martini:2011wp,Nieves:2011pp,Nieves:2011yp,Megias:2014qva,Gonzalez-Jimenez:2014eqa,Pandey:2014tza}. 
In particular, the Green's function Monte Carlo (GFMC) method~\cite{Carlson:2014vla} have been successfully applied to perform {\it first principle} calculations of the neutral-current response functions in the quasielastic (QE) region, up to moderate values of the momentum transfer~\cite{Lovato:2014eva,Lovato:2017cux}. GFMC results have unambiguously identified the role of nuclear correlations and meson-exchange currents in providing the most accurate description of lepton-nucleus scattering. Although QE processes dominate the total cross section for neutrino fluxes in the sub-GeV region, as in T2K~\cite{t2k_web} and MicroBooNE~\cite{microboone_web} experiments, pion-production constitutes an important background. A signal corresponding to a pion produced in the primary vertex and later absorbed in the nucleus could be misidentified with a QE event. Accurate predictions for inelastic channels are fundamental for experiments characterized by higher neutrino energies, such as MINER$\nu$A~\cite{minerva_web}, NO$\nu$A~\cite{nova_web}. and DUNE~\cite{dune_web}. Extending the applicability of GFMC to processes with energies higher than those corresponding to the QE kinematics poses nontrivial difficulties. The use of integral-transform techniques precludes a proper treatment of the energy dependence of the current operators. In addition, despite strategies to include the leading relativistic effects in the kinematics exist, the explicit inclusion of pions, needed for a proper description of the resonance region, is still in its infancy~\cite{Madeira:2018ykd}.

The framework based on the impulse approximation (IA) and realistic hole spectral-functions (SFs) is ideally suited to combine a realistic description of the initial target state --  as in the GFMC a realistic phenomenological Hamiltonian is employed  -- with a fully-relativistic interaction vertex and kinematics~\cite{Benhar:2006wy}. In its original formulation, this factorization scheme relies on the assumption that lepton-nucleus scattering reduces to the incoherent sum of elementary processes involving individual nucleons. Over the past few years, the IA was generalized to include the excitation of two particle-two hole final states induced by relativistic meson-exchange currents~\cite{Benhar:2015ula}. This extended factorization scheme (EFS) has been applied to calculate the electroweak inclusive cross sections of carbon and oxygen~\cite{Rocco:2015cil,Rocco:2018mwt}.

Early investigations of real-pion emission in inclusive $^{12}$C$(e,e')$ scattering were carried out within the IA in Refs.~\cite{Blomqvist:1977rv,Koch:1985qz,Chen:1988xq}. There, the elementary $\gamma^* N \rightarrow \pi N$ cross sections, generated from tree-diagram models consisting of the standard Born terms and the excitation of the $\Delta$(1232) resonance, were convolved with the nucleon momentum distributions. The two-nucleon mechanism $\gamma^* + NN \rightarrow \Delta N\rightarrow  NN$ was included~\cite{Chen:1988xq} with the parameters determined by fitting the total cross section data of $\gamma + d \rightarrow np$ reaction in the $\Delta$ excitation region. Medium effects on the $\Delta$ propagation, mainly due to the pion absorption within the nucleus~\cite{Lee:2002eq} were modeled using a $\Delta$-nucleus potential, which was phenomenologically determined within the isobar-hole model of $\pi$-nucleus scattering~\cite{Hirata:1977hg,Hirata:1978wp,Horikawa:1980cv}. The authors of Ref.~\cite{Szczerbinska:2006wk} improved upon the above procedure by considering the $\gamma^* N \rightarrow \pi N$ amplitudes generated from a dynamical model~\cite{Sato:1996gk,Sato:2000jf}, known as the Sato-Lee (SL) model, which provides a unified description of $\pi N \rightarrow \pi N$, $\gamma N \rightarrow \pi N$, and $N(e,e'\pi)N$ reactions up to the $\Delta$-excitation region. The correlated-basis function (CBF) hole-SF~\cite{Benhar:1989aw,Benhar:1994hw} was used to account for the nuclear correlations in the initial target state. Predictions were also made for neutrino-induced inclusive $^{12}C(\nu,\mu)$ cross sections using the extended SL model~\cite{Sato:2003rq}, which also contains axial current contributions fit to $N(\nu,\mu \pi)N$ reactions data. On the other hand, two-nucleon $\gamma^* + NN \rightarrow  NN$ mechanisms were not considered in that analysis.

In this work, we have implemented into the EFS the electroweak pion production amplitudes generated from the dynamical coupled-channel model~\cite{Kamano:2013iva,Nakamura:2015rta,Kamano:2016bgm} developed by the Argonne National Laboratory-Osaka University (ANL-Osaka) collaboration. The ANL-Osaka model is an extension of the SL model to include all important  meson-baryon channels  and all nucleon resonances up to invariant mass $W=$ 2 GeV. The parameters of the ANL-Osaka model are determined~\cite{Kamano:2013iva,Kamano:2016bgm} by fitting about 26,000 data points of the $\pi N, \gamma N\to \pi N, \eta N, K\Lambda, K\Sigma$ data from the channel thresholds to $W\le 2.1$~GeV. It had also been extended~\cite{Nakamura:2015rta} to describe the data of neutrino-induced $N(\nu,\mu\,\pi)N$ transitions. The resulting model contains about 20 nucleon resonances, which include all of the 4-star resonances listed by the Particle Data Group~\cite{PDG}. Clearly the use of the ANL-Osaka model makes this work significantly different from the above-mentioned references. It has to be noted that a more recent work~\cite{Vagnoni:2017hll}, which utilized the Paschos-Lalakuklich amplitudes~\cite{Paschos:2003qr,Lalakulich:2005cs,Lalakulich:2006sw}, only includes P$_{33}$(1232), D$_{13}$(1520), P$_{11}$(1440), and S$_{11}$(1535) resonances.

We have also significantly improved upon the treatment of medium effects in the $\Delta$-component of the two-body current, whose importance was established in the investigation of pion-nucleus reactions, as reviewed in Ref.\cite{Lee:2002eq}. On the same line as Ref.~\cite{Chen:1988xq}, we introduce in the $\Delta$ propagator of the two-body current a $\Delta$-nucleus potential. To account for the momentum-dependence of the medium effects, the latter is generated~\cite{Lee:1981st,Lee:1996mv} from a Bruckner-Hartree-Fock calculation based on a coupled-channel $NN\oplus N\Delta\oplus \pi NN$ model~\cite{Lee:1983xu,Lee:1984us,Lee:1985jq,Lee:1987hd}. 

To test the reliability of our approach, we first calculate the electron-$^{12}$C inclusive cross sections for a variety of kinematical setups, assessing the relative importance of the different reaction mechanisms. We also present results for neutrino and anti-neutrino-$^{12}$C scattering, induced by both neutral- and charged-current transitions. We refrain from presenting flux-folded calculations, since the latter require a more refined treatment of final-state interactions (FSI), for both two-nucleon knockout and pion-production processes. In particular, as far as the pion-production region is concerned, the processes in which the pion produced in the interaction vertex is absorbed in the nuclear medium should be accounted for before meaningful comparison with data are made. 

In Section~\ref{sec:eia} we report the expressions for the lepton-nucleus inclusive cross sections in terms of the relevant response functions. One- and two-body current reaction mechanisms are reviewed in Section~\ref{sec:obp} and~\ref{mec:sec}, while Section~\ref{pion:prod:sec} is devoted to pion-production processes. In Section~\ref{sec:res} we present our results on leptons scattering off $^{12}$C and in Section~\ref{sec:concl} we state our conclusions.

\section{Formulation of electroweak lepton-nucleus inclusive cross sections}
Let us consider a charge-changing process in which a neutrino ($\nu_\ell$) or an anti-neutrino ($\bar{\nu}_\ell$) with initial momentum $k^\mu=(E,\mathbf{k})$ scatters off a nuclear target, the final hadronic state being undetected. Denoting by $k^{\prime \mu}=(E^\prime,\mathbf{k}^\prime)$ the momentum of the outgoing lepton, the double-differential cross section in the Born approximation can be written as~\cite{Shen:2012xz,Benhar:2006nr}
\begin{equation}
\Big(\frac{d\sigma}{dE^\prime d\Omega^\prime}\Big)_{\nu_\ell/\bar{\nu}_\ell}= \frac{G_F^2  \cos^2\theta_c}{4\pi^2}\, k^\prime E^\prime\, L_{\mu \nu} W^{\mu \nu}\,.
\label{eq:xsec_def}
\end{equation}
We take $\cos\theta_c=0.97425$~\cite{Nakamura:2010zzi} and for the Fermi coupling constant we adopt the value $G_F = 1.1803 \times 10^{-5}\,\rm GeV^{-2}$, as from the analysis of $0^+ \to 0^+$ nuclear $\beta$-decays of Ref.~\cite{Herczeg:1999}, which accounts for the bulk of the inner radiative corrections~\cite{Nakamura:2002jg}.

The leptonic tensor is fully determined by the kinematics of the leptons in the initial and final states
\begin{equation}
L_{\mu \nu}  = \frac{1}{EE^\prime} (k_\mu k^\prime_\nu + k^\prime_\mu k_\nu - g_{\mu\nu}\, k \cdot k^\prime \pm i \epsilon_{\mu\rho\nu\sigma} k^\rho k^{\prime\, \sigma} )\, ,
\label{eq:lepton_def}
\end{equation}
where the $+$ $(-)$ sign is for $\nu_\ell$ ($\bar{\nu_\ell}$) initiated reactions. 
The hadronic tensor, containing all information on strong-interaction dynamics of the target nucleus, is defined in terms of the transition between the initial and final nuclear states $|\Psi_0\rangle$ and $|\Psi_f\rangle$, with energies $E_0$ and $E_f$. For spin-zero nuclei it can be cast in the form 
\begin{align}
W^{\mu\nu}= \sum_f \langle \Psi_0|j^{\mu \, \dagger}_{CC}|\Psi_f\rangle \langle \Psi_f| j^\nu_{CC} |\Psi_0 \rangle \delta (E_0+\omega -E_f)\, ,
\label{eq:had_tens}
\end{align}
where the charged-current operator is the sum of a vector and axial component $j^\mu_{CC} = j^\mu_{V}+j^\mu_{A}$.

Taking the three-momentum transfer along the $z$ axis and the total three-momentum in the $x-z$ plane
\begin{align}
q&=k-k^\prime=(\omega, {\bf q})\, , \quad {\bf q}= (0,0,q_z)\nonumber\\ 
Q&=k+k^\prime=(\Omega, {\bf Q}) \, , \quad {\bf Q}= (Q_x,0,Q_z)\, ,
\end{align}
performing the Lorentz contraction in Eq.~\eqref{eq:xsec_def} yields
\begin{align}
\Big(\frac{d\sigma}{dE^\prime d\Omega^\prime }\Big)_{\nu/\bar{\nu}}&=\frac{G_F^2  \cos^2\theta_c}{4\pi^2}\frac{k^\prime}{2 E_\nu}\left[\hat{L}_{CC}R_{CC}+2\hat{L}_{CL}R_{CL}\right. \nonumber\\
&\left.+\hat{L}_{LL}R_{LL}+\hat{L}_{T}R_{T}\pm 2 \hat{L}_{T^\prime}R_{T^\prime}\right]\ ,
\label{eq:cross_sec}
\end{align}
where the kinematical factors are given by
\begin{align}
\hat{L}_{CC}&={\Omega}^2-q_z^2 -m_\ell^2\nonumber\\
\hat{L}_{CL}&=(-\Omega Q_z+\omega q_z)\nonumber\\
\hat{L}_{LL}&= {Q_z}^2 - \omega^2 + m_\ell^2\nonumber\\
\hat{L}_{T}&= \frac{{Q_x}^2}{2}- q^2 + m_\ell^2 \nonumber\\
\hat{L}_{T^\prime}&= \Omega q_z- \omega Q_z\, ,
\end{align}
and $m_\ell^2=k^{\prime\, 2}$ is the mass of the outgoing lepton. 
The five electroweak response functions are expressed in terms of the hadron tensor components as 
\begin{align}
R_{CC}&=W^{00}\nonumber\\
R_{CL}&=-\frac{1}{2}(W^{03}+W^{30})\nonumber\\
R_{LL}&=W^{33}\nonumber\\
R_T&=W^{11}+W^{22}\nonumber\\
R_{T^\prime}&=-\frac{i}{2}(W^{12}-W^{21})\,.
\label{eq:response_w}
\end{align}

Note that the inclusive cross section of an electron scattering off a nucleus in the one-photon exchange approximation can be written in a similar fashion as in Eq.~\eqref{eq:xsec_def}, provided that $G^2/4\pi^2$ is replaced by $2\alpha^2/q^4$, where $\alpha\simeq1/137$ is the fine structure constant, and the contribution proportional to the Levi Civita tensor is dropped from the leptonic tensor of Eq.~\eqref{eq:lepton_def}. Hence, the double-differential cross section for this process reads 
\begin{align}
\Big(\frac{d\sigma}{dE^\prime d\Omega^\prime}\Big)_{e}&=\left( \frac{d \sigma}{d\Omega^\prime} \right)_{\rm{M}} \Big[ \hat{A}_{CC}\,  R_{CC} \nonumber \\
& + \hat{A}_T\,  R_T \Big] \ ,
\label{eq:x:sec}
\end{align}
where 
\begin{align}
\hat{A}_{CC} = \Big( \frac{q^2}{q_z^2}\Big)^2  \ \ \ , \ \ \ \hat{A}_T = -\frac{1}{2}\frac{q^2}{q_z^2}+\tan^2\frac{\theta}{2}\,, 
\end{align}
$\theta$ being the lepton scattering angle and
\begin{align}
\label{Mott}
\left( \frac{d \sigma}{d \Omega^\prime} \right)_{\rm{M}}= \left[ \frac{\alpha \cos(\theta/2)}{2 E^\prime\sin^2(\theta/2) }\right]^2
\end{align} 
is the Mott cross section. The electromagnetic responses of Eq.~\eqref{eq:x:sec} are written in terms of the hadron tensor components as in Eq.~\eqref{eq:response_w}, provided that  $j^\mu_{CC}$ is replaced by the electromagnetic current $j^\mu_{EM}$, which is related to $j^\mu_{V}$ through the conserved vector current (CVC) hypothesis. Because of their striking similarities and common ingredients, it is evident that a prerequisite for any reliable model of neutrino-nucleus scattering is its capability of accurately describing the large body of measured electron-scattering cross sections~\cite{Benhar:2006er}.

\section{Extended Impulse approximation}
\label{sec:eia}
The initial state of the target nucleus appearing in Eq.~\eqref{eq:had_tens} does not depend on momentum transfer and can be safely treated within nuclear many-body theory (NMBT) regardless the kinematics of the scattering. Within this scheme, the nucleus is viewed as a collection of $A$ pointlike protons and neutrons, whose dynamics are described by the nonrelativistic Hamiltonian
\begin{equation}
H=\sum_{i}\frac{{\bf p}_i^2}{2m_N}+\sum_{j>i} v_{ij}+ \sum_{k>j>i}V_{ijk}\ .
\label{NMBT:ham}
\end{equation}
In the above equation, ${\bf p}_i$ is the momentum of the $i$-th nucleon of mass $m_N$, while the potentials $v_{ij}$ and $V_{ijk}$ model the nucleon-nucleon (NN) and three-nucleon (3N) interactions, respectively.  Up to moderate values of the momentum transfer, typically $|\mathbf{q}| \lesssim 500$ MeV, NMBT can be applied to compute the response functions of $A\leq 12$ nuclei using initial- and final-state nuclear wave functions derived from the Hamiltonian of Eq.~\eqref{NMBT:ham}. In particular, virtually-exact Green's function Monte Carlo (GFMC) calculations have shown that the strength and energy-dependence of two-nucleon processes induced by correlation effects and interaction currents are crucial in providing the most accurate description of electron- and neutrino-nucleus scattering in the quasielastic regime~\cite{Lovato:2016gkq,Lovato:2017cux}.

At large values of energy and momentum transfer, a calculation of the hadron tensor solely based on NMBT is no longer reliable. In this regime, the final state includes at least one particle carrying a large momentum $\sim \mathbf{q}$, and fully-relativistic expressions of the transition currents need to be retained. The impulse approximation (IA) scheme allows one to circumvent the difficulties associated with the relativistic treatment of $|\Psi_f\rangle$ and of the current operator, while at the same time preserving essential features (such as correlations) inherent to a realistic description of nuclear dynamics. 

\subsection{One-body current processes}
\label{sec:obp}
The IA scheme is based on the tenet that for $|\mathbf{q}| \gg 1/d$, $d$ being the average nucleon-nucleon separation distance in the target, the struck nucleon is largely decoupled from the spectator $(A-1)$ particles~\cite{Benhar:2006wy, Benhar:2015wva}. Within the original implementation of the IA, the nuclear current operator reduces to a sum of one-body terms
\begin{equation}
j^\mu=\sum_i j^\mu_i
\end{equation}
and the nuclear final state factorizes as
\begin{align}  
|\psi_f^A \rangle \rightarrow |p\rangle \otimes |\psi_f^{A-1}\rangle\, .
\label{eq:factorization} 
\end{align}
In the above equation $|p\rangle$ denotes the final-state nucleon with momentum ${\bf p}$ and energy $e({\bf p})=\sqrt{{\bf p}^2+m_N^2} $, while $|\psi_f^{A-1}\rangle$ describes the \hbox{$(A-1)$-body} spectator system. Its energy and recoiling momentum are fixed by energy and momentum conservation
\begin{align}
E_f^{A-1}=\omega +E_0-e({\bf p})\, ,\quad {\bf P}^{A-1}_f={\bf q}-{\bf p}\, .
\end{align}
Employing the factorized expression of the nuclear final state in Eq.~\eqref{eq:had_tens} and inserting a single-nucleon completeness relation, the incoherent contribution to the one-body (1b) hadron tensor is given by
\begin{align}
W^{\mu\nu}_{\rm 1b}({\bf q},\omega)=& \int \frac{d^3k}{(2\pi)^3} dE P_h({\bf k},E)\frac{m_N^2}{e({\bf k})e({\bf k+q})}\nonumber\\
&\times  \sum_{i}\, \langle k | {j_{i}^\mu}^\dagger |k+q \rangle \langle k+q |  j_{i}^\nu | k \rangle\nonumber\\
& \times \delta(\omega+E-e(\mathbf{k+q}))\,,
\label{had:tens}
\end{align}
The factors $m_N/e({\bf k})$ and $m_N/e({\bf k+q})$ are included to account for the implicit covariant normalization of the four-spinors of the initial and final nucleons in the matrix elements of the relativistic current. The hole spectral function
\begin{align}
P_h(\mathbf{k},E)&=\sum_f |\langle \psi_0^A| [|k\rangle \otimes |\psi_f^{A-1}\rangle]|^2\nonumber\\
&\times\delta(E+E_{f}^{A-1}-E^A_0)
\label{pke:hole}
\end{align}
provides the probability distribution of removing a nucleon with momentum ${\bf k}$ from the target nucleus, leaving the residual $(A-1)$-nucleon system with an excitation energy $E$.
Note that in Eq.~\eqref{had:tens} we neglected Coulomb interactions and the other (small) isospin-breaking terms and made the assumption, largely justified in the case of symmetric isospin zero (T=0)
nuclei, that the proton and neutron spectral functions are identical.

Within the correlated-basis function theory (CBF), the hole SF of finite nuclei is expressed as a sum of two contributions~\cite{Benhar:1994hw}, displaying distinctly different energy and momentum dependences
\begin{align}
P_h({\bf k},E)= P^{ 1h}_h({\bf k},E) + P_h^{\rm corr}({\bf k},E)\, .
\label{eq:fullPh}
\end{align}
The one-hole term, corresponding to bound $A-1$ states, is obtained from a modified mean-field scheme
\begin{align}
\label{Pke:MF}
P^{1h}_h({\bf k},E) = \sum_{\alpha\in \{{\rm F}\}} Z_\alpha |\phi_\alpha({\bf k})|^2 F_\alpha(E-e_\alpha) \ , 
\end{align}
where the sum runs over all occupied single-particle nuclear states, labeled by the index $\alpha$, and $\phi_\alpha({\bf k})$ is the Fourier transform of the shell-model orbital with energy $e_\alpha$. The {\it spectroscopic} factor $Z_\alpha < 1$ and the function $F_\alpha(E-e_\alpha)$, describing the energy width of the state $\alpha$, account for the effects of residual interactions that are not included in the mean-field picture. In the absence of the latter, $Z_\alpha \to 1$  and $F_\alpha(E-e_\alpha) \to \delta_\alpha(E-e_\alpha)$. The spectroscopic factors and the widths of the $s$ and $p$ states of $^{12}$C used in this work are from the analysis of $(e,e^\prime p)$ data carried out in Refs.~\cite{Mougey:1976sc,Turck:1981,Dutta:2003yt}.

The correlated part of the SF for finite nuclei $P_h^{\rm corr}({\bf k},E)$ corresponds to unbound $|\psi_f^{A-1}\rangle$ states in Eq.~\eqref{pke:hole}, in which at least one of the spectators is in the continuum. It is obtained through the local density approximation (LDA) procedure
\begin{align}
P^{ \rm corr}_h({\bf k},E) = \int d^3R \  \rho_A({\bf R}) P^{\rm corr}_{h,\,NM}({\bf k},E; \rho_A({\bf R}))\,,
\label{Pke:corr}
\end{align}
In the above equation, $\rho_A(\mathbf{R})$ is the nuclear density distribution of the nucleus and $P^{\rm corr}_{h\,,NM}({\bf k},E; \rho)$ is the correlation component of the SF of isospin-symmetric nuclear matter at density $\rho$, which vanishes if nuclear correlations are not accounted for. The use of the LDA to account for $P^{\rm corr}_h({\bf k},E)$ is justified by the fact that to a remarkably large extent short-range nuclear dynamics is unaffected by surface and shell effects. The energy-dependence exhibited by $P^{\rm corr}_h({\bf k},E)$, showing a widespread background extending up to large values of both $k$ and  $E$, is completely different from that of $P^{ 1h}_h({\bf k},E)$. For $k>p_F$, $P^{\rm corr}_h({\bf k},E)$ coincides with $P_h({\bf k},E)$ and its integral over the energy gives the so-called continuous part of the momentum distribution.

\begin{figure}[]
\centering
\includegraphics[width=\columnwidth]{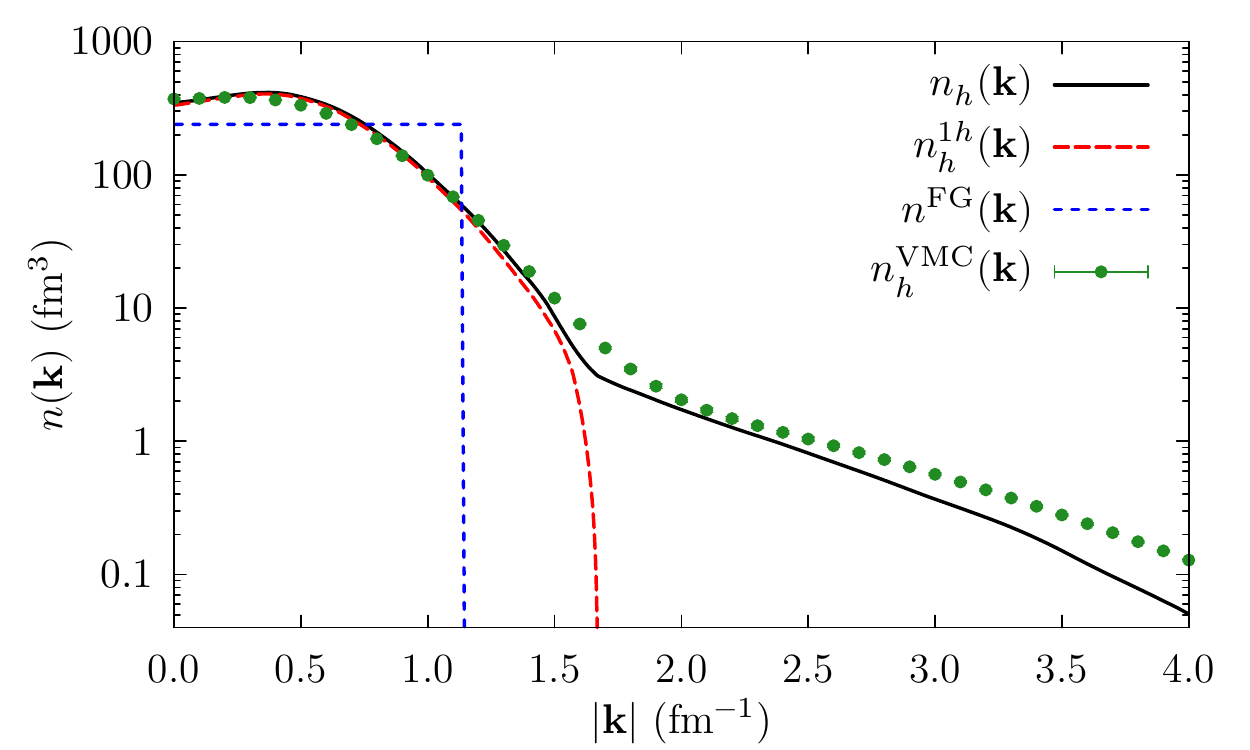}
\caption{Momentum distributions associated with the hole SF ($n_h({\bf k})$), the mean-field component of the hole SF ($n_h^{1h}({\bf k})$), the free Fermi gas at $k_F=225$ MeV ($n^{\rm FG}({\bf k})$), and the VMC results of~\cite{theory_web} ($n_h^{\rm VMC}({\bf k})$).}
\label{fig:nk}
\end{figure}

The distinct momentum dependences of the one-hole and the correlated part of the hole SF can be appreciated by comparing the momentum distributions corresponding to $P_h({\bf k},E)$ and $P^{ 1h}_h({\bf k},E)$, displayed in Fig.~\ref{fig:nk}. In this figure we also show the free Fermi gas momentum distribution for $k_F=225$ MeV and the one computed within variational Monte Carlo (VMC) using a Hamiltonian comprised of the Argonne $v_{18}$~\cite{Wiringa:1994wb} and the Urbana X~\cite{Carlson:2014vla} potentials. It is clear that the correlation component  enhances the high-momentum tail of the hole SF bringing the corresponding momentum distribution in is good agreement with the VMC results~\cite{theory_web}. On the other hand, the differences with the free Fermi gas approximation are striking: $n^{\rm FG}({\bf k})$ is flat for $|{\bf k}| < k_F$ and vanishes for $|{\bf k}|> k_F$. 

In the kinematical region in which the interactions between the struck particle and the spectator system can not be neglected, the IA results have to be modified to include the effect of FSI. Following Ref.~\cite{Ankowski:2014yfa}, we consider the real part of the optical potential $U$ derived from the Dirac phenomenological fit of Ref.~\cite{Cooper:1993nx} to describe the propagation of the knocked-out particle in the mean-field generated by the spectator system. This potential, given as a function of the kinetic energy of the nucleon $t_{kin}(\mathbf{p})=\sqrt{{\bf p}^2+m^2}-m$, modifies the energy spectrum of the struck nucleon as
\begin{equation}
\tilde{e}({\bf k+q})=e({\bf k+q})+ U\left(t_{\rm kin}({\bf k+q})\right)\, .
\end{equation}
The multiple scatterings that the struck particle undergoes during its propagation through the nuclear medium are taken into account through a convolution scheme. The IA responses are folded with the function $f_\mathbf{k+q}$, normalized as 
\begin{equation}
\int_{-\infty}^{+\infty} d\omega f_{\bf k+q}(\omega) = 1\ .
\label{fold:func}
\end{equation}
The one-body hadron tensor then reads
\begin{align}
W^{\mu\nu}_{\rm 1b}({\bf q},\omega)=& \int \frac{d^3k}{(2\pi)^3} dE P_h({\bf k},E)  \int d\omega^\prime\,f_{\bf k+q}(\omega-\omega^\prime)\nonumber\\
&\times  \frac{m_N^2}{e({\bf k})e({\bf k+q})} \sum_{i}\, \langle k | {j_{i}^\mu}^\dagger |k+q \rangle \langle k+q |  j_{i}^\nu | k \rangle\nonumber\\
& \times\delta(\omega^\prime +E-\tilde{e}({\bf k+q})) \theta(|{\bf k+q}|-p_F)\,,
\label{dens:resp:folding}
\end{align}
The folding function is computed within a generalization of the Glauber theory~\cite{Benhar:1991af}
\begin{align}
f_{\bf p}(\omega)=&\ \delta(\omega)\sqrt{T_{\bf p}}+\int \frac{dt}{2\pi} e^{i\omega t}\left[\bar{U}^{FSI}_{\bf p}(t)-\sqrt{T_{\bf p}}\ \right]\nonumber\\
=&\ \delta(\omega)\sqrt{T_{\bf p}}+(1-\sqrt{T_{\bf p}})F_{\bf p}(\omega)\, ,
\label{fold:func:expl}
\end{align}
Full expressions for the nuclear transparency $T_{\bf p}$ and for the finite width function $F_{\bf p}(\omega)$ can be found in~\cite{Benhar:2006wy,Benhar:2013dq}. 

The one-body CC operator is the sum of a vector and axial component 
\begin{align}
j^\mu_{CC}&=j^\mu_{V}+j^\mu_{A}\nonumber\\
j^\mu_{V}&={\mathcal F}_1\gamma^\mu + i \sigma^{\mu\nu}q_\nu \frac{{\mathcal F}_2}{2m_N}\nonumber\\
j^\mu_{A}&=-\gamma^\mu \gamma_5 {\mathcal F}_A-q^\mu \gamma_5 \frac{{\mathcal F}_p}{m_N}, 
\end{align}
where 
\begin{align}
{\mathcal F}_1=& F_{1}^V \tau_\pm \nonumber\\
{\mathcal F}_2=& F_{2}^V \tau_\pm\, 
\end{align}
and $\tau_\pm= (\tau_x \pm i \tau_y)/2$ is the isospin raising/lowering operator. The Dirac and Pauli form factors defining $F_{1,2}^V=F_{1,2}^p - F_{1,2}^n$ are usually written in terms of the Sachs form factors as
\begin{align}
F_1^{p,n}=&\frac{G_E^{p,n}+\tau G_M^{p,n}}{1+\tau}\nonumber\\
F_2^{p,n}=&\frac{G_M^{p,n}-G_E^{p,n}}{1+\tau}
\end{align}
with $\tau=-q^2/4m_N^2$. The axial term of the CC can be cast in the form
\begin{align}
{\mathcal F}_A=& F_{A} \tau_\pm \nonumber\\
{\mathcal F}_P=& F_{P} \tau_\pm\,.
\end{align}
We employ the standard dipole parametrization for the axial form factor
\begin{align}
 F_A &=\frac{g_A}{( 1- q^2/ m_A^2 )^2}\ ,
\end{align}
where the nucleon axial-vector coupling constant is taken to be $g_A=1.2694$~\cite{PDG} and the axial mass $m_A=1.049$ GeV. Uncertainties in the $Q^2$ dependence of the axial form factor impact neutrino-nucleus cross-section predictions. In this regard, the dipole parametrization has been the subject of intense debate: dedicated lattice-QCD calculations of $G_A(Q^2)$ have been carried out~\cite{Rajan:2017lxk} and an alternative ``z-expansion'' analyses~\cite{Meyer:2016oeg} has been recently proposed.

Partially Conserved Axial Current (PCAC) arguments connect the pseudo-scalar form factor to the axial one
\begin{align}
F_P=\frac{2m_N^2}{(m_\pi^2-q^2)}F_A\, ,
\end{align}
$m_\pi$ being the pion mass. While $F_P$ can be neglected when considering $\nu_e$, $\nu_\mu$-induced processes, its contribution cannot be ignored for a heavy $\tau$ lepton production~\cite{Sobczyk:2019urm} and in the analysis of muon-capture processes~\cite{Lovato:2019fiw}. 

The conserved-vector-current (CVC) hypothesis allows one to relate the vector component of the CC current to the EM: $ j_{\rm EM}^\mu=j_{V}^\mu$,
provided that
\begin{align}
{\mathcal F}_1 & = \frac{1}{2}[F_1^S + F_1^V\tau_z]\nonumber\\
{\mathcal F}_2 & = \frac{1}{2}[F_2^S+F_2^V\tau_z]
\end{align}
where $F_{1,2}^S=F_{1,2}^p + F_{1,2}^n$ is the single-nucleon isoscalar form factor.

\subsection{Inclusion of two-body currents}
\label{mec:sec}
In the last few years, the IA scheme has been generalized to include meson-exchange currents, which naturally arise from the dynamics of the constituent nucleons. For instance, the gauge invariance of the theory imposes that the electromagnetic charge and current operators satisfy the continuity equation $\mathbf{q} \cdot \mathbf{j}_{\rm EM}=[H,j_{EM}^0]$. Since the two- and three-nucleon potentials of Eq.~\eqref{NMBT:ham} do not commute with the charge operator $j^\mu$ must comprise two- and three-nucleon contributions. Neglecting the latter, which have numerically proven to be very small in $A=3$ observables~\cite{Marcucci:2005zc}, we can write the CC and EM currents as 
\begin{equation}
j^\mu=\sum_i j^\mu_i + \sum_{i<j} j^\mu_{ij}\, .
\end{equation}

In Refs.~\cite{Benhar:2015ula,Rocco:2015cil,Rocco:2018mwt} the factorization ansatz of Eq.~\eqref{eq:factorization} has been extended to treat the amplitudes involving two-nucleon currents consistently with the correlation component of the hole SF
\begin{equation}
|\psi_f^A \rangle \rightarrow |p p^\prime \rangle_a \otimes |\psi_f^{A-2}\rangle\, .
\end{equation}
where $|p\,p^\prime\rangle_a=|p\,p^\prime\rangle-|p^\prime\,p\rangle$. In infinite isospin-symmetric nuclear matter, the pure two-body current component of the hadron tensor turns out to be~\cite{Benhar:2015ula}
\begin{align}
&W^{\mu\nu}_{\rm 2b}({\bf q},\omega)=\frac{V}{4} \int dE \frac{d^3k}{(2\pi)^3}  \frac{d^3k^\prime}{(2\pi)^3}\frac{d^3p}{(2\pi)^3}
\frac{m_N^4}{e({\bf k})e({\bf k^\prime})e({\bf p})e({\bf p^\prime})} \nonumber\\
 &\qquad \times  P_h^{\rm NM}({\bf k},{\bf k}^\prime,E) 2\sum_{ij}\, \langle k\, k^\prime | {j_{ij}^\mu}^\dagger |p\,p^\prime\rangle_a \langle p\,p^\prime |  j_{ij}^\nu | k\, k^\prime \rangle\nonumber\\
 &\qquad \times \delta(\omega+E-e(\mathbf{p})-e(\mathbf{p}^\prime))\, .
\end{align}
In the above equation, the normalization volume for the nuclear wave functions $V=\rho / A$ with $\rho=3\pi^2 k_F^3/2$ depends on the Fermi momentum of the nucleus, which for $^{12}$C we take to be $k_F=225$ MeV. The factor $1/4$ accounts for the sum over indistinguishable pairs of particles, while the factor $2$ arises from the fact that, renaming the dummy indexes, the product of the two direct terms is equal to the one of the two exchange terms \cite{Dekker:1991ph}. In principle, the calculation of $W^{\mu\nu}_{\rm 2b}({\bf q},\omega)$ requires the knowledge of the two-nucleon hole spectral function of infinite nuclear matter $P_h^{\rm NM}({\bf k},{\bf k}^\prime,E)$. Within the CBF theory, it has been shown that, in absence of long-range correlations, the two-body momentum distribution factorizes as
\begin{equation}
\int dE P_h^{\rm NM}({\bf k},{\bf k}^\prime,E) = n({\bf k},{\bf k}^\prime)=n({\bf k})n({\bf k}^\prime)+ {\cal O}\bigg(\frac{1}{A}\bigg)\, .
\label{eq:nk1k2}
\end{equation}
Hence, the two-body current component of the hadron tensor can be expressed as
\begin{align}
W^{\mu\nu}_{\rm 2b}({\bf q},\omega)&= \frac{V}{2}  \int d\tilde{E} \frac{d^3k}{(2\pi)^3} d\tilde{E}^\prime \frac{d^3k^\prime}{(2\pi)^3}\frac{d^3p}{(2\pi)^3}
 \nonumber\\
 &\times \frac{m_N^4}{e({\bf k})e({\bf k^\prime})e({\bf p})e({\bf p^\prime})} P_h^{\rm NM}({\bf k},\tilde{E})P_h^{\rm NM}({\bf k}^\prime,\tilde{E}^\prime)\nonumber\\
&\times  \sum_{ij}\, \langle k\, k^\prime | {j_{ij}^\mu}^\dagger |p\,p^\prime \rangle_a \langle p\,p^\prime |  j_{ij}^\nu | k\, k^\prime \rangle\nonumber\\
& \times \delta(\omega+\tilde{E}+\tilde{E}^\prime-e(\mathbf{p})-e(\mathbf{p}^\prime))\, .
\label{had:tens2}
\end{align}
In order to treat atomic nuclei, following Ref.~\cite{Rocco:2018mwt}, we replace the hole SF of infinite matter with the one of $^{12}$C
\begin{equation}
P_h^{\rm NM}({\bf k},E) \to  \frac{k_F^3}{6\pi^2} P_h({\bf k},E)
\end{equation}
where $P_h({\bf k},E)$ is computed as in Eq.~\eqref{eq:fullPh}. 

It has been argued that the strong isospin dependence of the two-nucleon momentum distribution, supported by experimental data, persist for nuclei even larger than $^{12}$C~\cite{Wiringa:2013ala,Weiss:2015mba,Weiss:2018tbu,Lonardoni:2018sqo}, hence questioning the regime of applicability of Eq.~\eqref{eq:nk1k2}. A viable strategy to gauge the limitations of the factorization of the two-body momentum distribution consists in approximating the latter with the so-called two-body decay function~\cite{Frankfurt:1988nt}
\begin{equation}
P_h^{\rm NM}({\bf k},{\bf k}^\prime,E) \to n({\bf k},{\bf k}^\prime) \delta(E-\bar{E}_f^{A-2})\, ,
\end{equation}
$\bar{E}_f^{A-2}$ being the average energy of the $A-2$ spectator system, and use variational Monte Carlo results for $n({\bf k},{\bf k}^\prime)$. Explorative calculations in this directions are ongoing and will be the subject of a dedicated work. In this regard, it has to be noted that in this work the interference between one- and two-body currents is disregarded. While in the two-nucleon knockout final states this contribution is relatively small~\cite{Benhar:2015ula,Rocco:2015cil}, CBF calculations in infinite nuclear matter suggest that nuclear tensor correlations strongly enhance the interference terms for final states associated with single-nucleon knock out processes~\cite{Fabrocini:1996bu}. This is compatible with the Green's function Monte Carlo results for the electromagnetic~\cite{Lovato:2016gkq} and neutral-current response functions~\cite{Lovato:2017cux}, in which the interference between one- and two-body currents dominate the total two-body current contribution, significantly enhancing the quasielastic peak region. 

\begin{figure}[]
\centering
\includegraphics[width=\columnwidth]{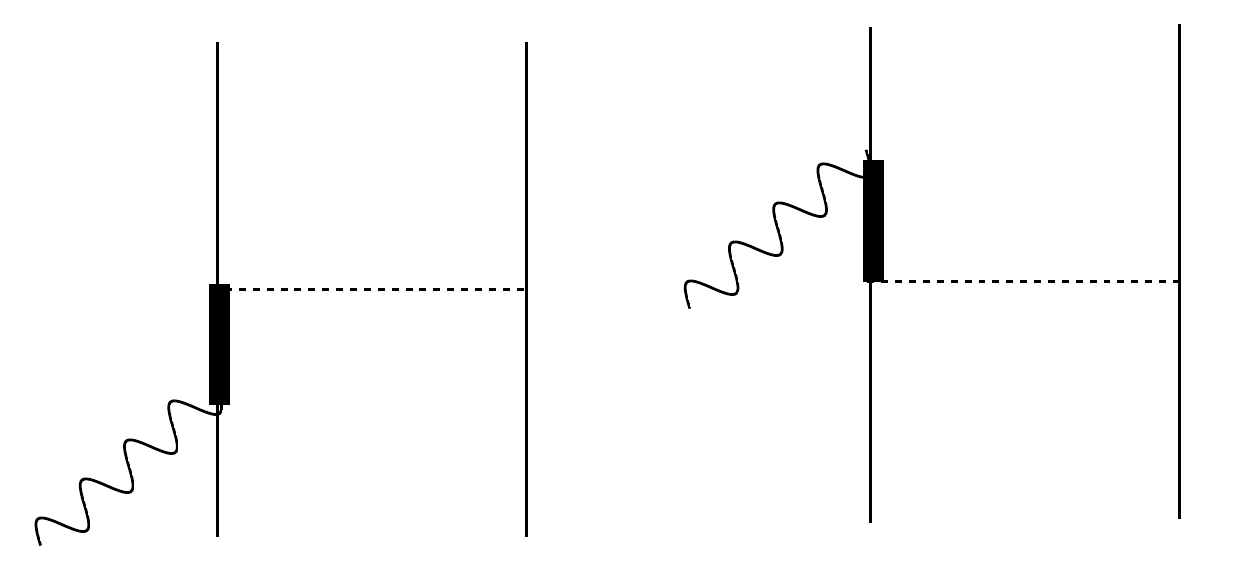}
\caption{Feynman diagrams describing two-body currents contributions associated to $\Delta$-excitations processes. Solid, thick, and dashed lines correspond to nucleons, deltas, pions, respectively. The wavy line represents the vector boson.}
\label{mec:diag}
\end{figure}

Analogously to the one-body case, the two-body CC operator is the sum of a vector and axial component. We use the  expressions derived in Ref.~\cite{Simo:2016ikv} by coupling the pion-production amplitudes of Ref.~\cite{Hernandez:2007qq} to a second nucleonic line. They can be traced back to four distinct interaction mechanisms, namely the pion in flight, seagull, pion-pole, and $\Delta$ excitations
\begin{align}
j^\mu_{\rm CC}= (j^\mu_{\rm pif})_{\rm CC}+(j^\mu_{\rm sea})_{\rm CC}+ (j^\mu_{\rm pole})_{\rm CC}+ (j^\mu_{\Delta})_{\rm CC}\, ,
\label{eq:jijCC}
\end{align}
The corresponding EM currents are obtained from the vector components of the $j^\mu_{\rm CC}$ using CVC hypothesis. Detailed expressions for the first four terms of Eq.~\eqref{eq:jijCC} can be found in~Refs.~\cite{Simo:2016ikv,Rocco:2018mwt}. Here we only focus on the diagrams reported in Fig.~\ref{mec:diag} (and the corresponding two in which particles 1 and 2 are interchanged), which are associated with  two-body current terms involving a $\Delta$-resonance in the intermediate state. Because of the purely transverse nature of this current, the form of its vector component is not subject to current-conservation constraints and its expression is largely model dependent. We adopted the parametrization of Ref.~\cite{Hernandez:2007qq}
\begin{align}
(j^\mu_\Delta)_{\rm CC}&=\frac{3}{2}\frac{f_{\pi NN} f^\ast}{m^2_\pi} \bigg\{ \Pi(k_2)_{(2)}
\Big[ \Big( -\frac{2}{3}\tau^{(2)}+\frac{I_V}{3}\Big)_{\pm} \nonumber\\
&\times F_{\pi NN}(k_2) F_{\pi N \Delta} (k_2) (j^\mu_{a})_{(1)}-\Big(\frac{2}{3}\tau^{(2)}+\frac{I_V}{3}\Big)_{\pm} \nonumber\\
&\times F_{\pi NN}(k_2) F_{\pi N \Delta} (k_2) (j^\mu_{b})_{(1)}\Big]+(1\leftrightarrow 2)\bigg\}
\label{delta:curr}
\end{align}
where $k_2$ is the momentum of the $\pi$ exchanged in the two depicted diagrams, $f^\ast$=2.14 and 
\begin{equation}
F_{\pi N \Delta}(k)=\frac{\Lambda^2_{\pi N\Delta}}{\Lambda^2_{\pi N\Delta}-k^2}\ ,
\end{equation}
with $\Lambda_{\pi N\Delta}=1150$ MeV. In the above equation, $j^\mu_a$ and $j^\mu_b$ denote the $N\rightarrow \Delta$ transition vertices of the left and right diagrams, respectively. They are expressed as 
\begin{align}
j^\mu_a&=(j^\mu_a)_V+(j^\mu_a)_A\ ,\nonumber\\
(j^\mu_a)_V&=\frac{C_3^V}{m_N}\Big[k_2^\alpha G_{\alpha\beta}(k+q)\Big(g^{\beta\mu}\slashed{q}-q^\beta\gamma^\mu\Big)\Big]\gamma_5\ ,\nonumber\\
(j^\mu_a)_A&=C_5^A \Big[k_2^\alpha G_{\alpha\beta}(k+q) g^{\beta\mu}\Big]
\end{align}
and
\begin{align}
j^\mu_b&=(j^\mu_b)_V+(j^\mu_b)_A\ ,\nonumber\\
(j^\mu_b)_V&=\frac{C_3^V}{m_N}\gamma_5\Big[\Big(g^{\alpha\mu}\slashed{q}-q^\alpha\gamma^\mu\Big)G_{\alpha\beta}(p-q)k_2^\beta\Big],\nonumber\\
(j^\mu_b)_A&=C_5^A \Big[ g^{\alpha\mu} G_{\alpha\beta}(p-q) k_2^\beta\Big]\, .
\end{align}
The Rarita-Schwinger propagator
\begin{align}
G^{\alpha\beta}(p_\Delta)=\frac{P^{\alpha\beta}(p_\Delta)}{p^2_\Delta-M_\Delta^2}
\label{eq:free_delta}
\end{align}
is proportional to the spin 3/2 projection operator 
\begin{align}
P^{\alpha\beta}(p_\Delta)&=(\slashed{p}_\Delta+M_\Delta)\Big[ g^\alpha\beta -\frac{1}{3}\gamma^\alpha\gamma^\beta  -\frac{2}{3}\frac{p_\Delta^\alpha p_\Delta^\beta}{M_\Delta^2}\nonumber\\
&+\frac{1}{3}\frac{p_\Delta^\alpha \gamma^\beta - p_\Delta^\beta \gamma^\alpha}{M_\Delta}\Big]\ .
\end{align}
The possible decay of the $\Delta$ into a physical $\pi N$ state is accounted for by replacing the real resonance mass $M_\Delta$=1232 MeV entering the free propagator of Eq.~\eqref{eq:free_delta} by $M_\Delta - i \Gamma(p_\Delta)/2$~\cite{Dekker:1994yc,DePace:2003spn}. The energy-dependent decay width $\Gamma(p_\Delta)/2$, 
effectively describing the allowed phase space for the pion produced in the decay, is given by
\begin{align}
\Gamma(p_\Delta)&=-2{\rm Im}\big[ \Sigma_{\pi N}(p_\Delta) \big]\nonumber\\
&=\frac{(4 f_{\pi N \Delta})^2}{12\pi m_\pi^2} \frac{|\mathbf{k}|^3}{\sqrt{s}} (m_N + E_k) R(\mathbf{r}^2)\, .
\label{eq:decay_width}
\end{align}
where $\Sigma_{\pi N}(p_\Delta) $ is the $\Delta$ self-energy in vacuum.
In the above equation, $(4 f_{\pi N \Delta})^2/(4\pi)=0.38$, $s=p_\Delta^2$ is the invariant mass, $\mathbf{k}$ is the decay three-momentum
in the $\pi N$ center of mass frame, such that
\begin{equation}
|\mathbf{k}|^2=\frac{1}{4s}[s-(m_N+m_\pi)^2][s-(m_N-m_\pi)^2]\,
\end{equation} 
and $E_k=\sqrt{m_N^2 + \mathbf{k}^2}$ is the associated energy. The additional factor
\begin{equation}
R(\mathbf{r}^2)=\left(\frac{\Lambda_R^2}{\Lambda_R^2-\mathbf{r}^2}\right)
\end{equation}
depending on the $\pi N$ three-momentum $\mathbf{r}$, with $\mathbf{r}^2=(E_k - \sqrt{m_\pi^2 + \mathbf{k}^2})^2-4\mathbf{k}^2$ and $\Lambda_R^2=0.95\, m_N^2$,
is introduced to improve the description of the experimental phase-shift $\delta_{33}$~\cite{Dekker:1994yc}.

\begin{figure}[h]
\centering
\includegraphics[width=\columnwidth]{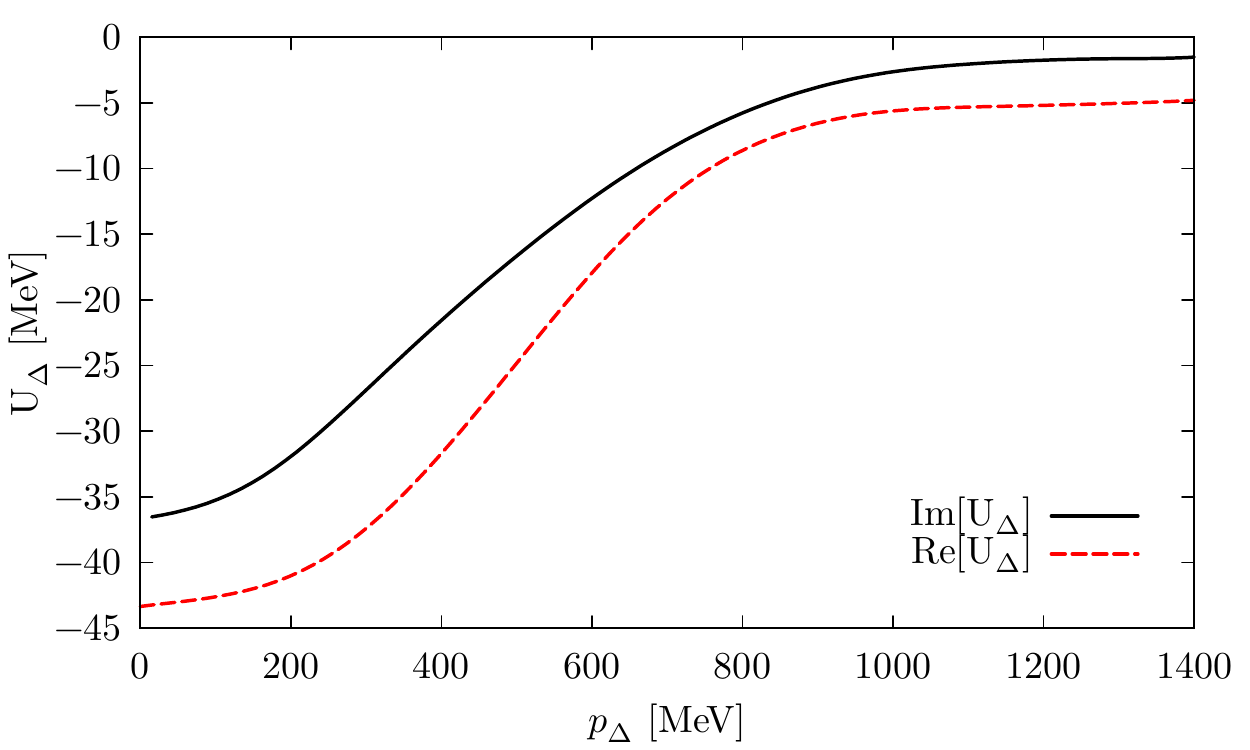}
\caption{Real and imaginary part of the $\Delta$ potential in nuclear matter at saturation density $\rho=0.16$ fm$^{-3}$.}
\label{fig:ud}
\end{figure}

To avoid potential double-counting with real-pion emission, the Authors of Refs.~\cite{DePace:2003spn,Simo:2016ikv,Butkevich:2017mnc,Rocco:2018mwt} adopted the prescription of retaining only the real part of the $\Delta$ propagator when computing two-body current processes. In the present work, we keep both the real and imaginary part of the $\Delta$-propagator and we introduce a phenomenological potential to model the $\Delta$ decay in the nucleus. 
Among the medium effects -- a detailed discussion of which can be found in Refs.~\cite{Hirata:1977hg,Hirata:1978wp,Horikawa:1980cv,Lee:2002eq,Lee:1981st,Lee:1996mv} -- the most important one is due to the annihilation of the $\Delta$ via $\Delta N \rightarrow NN$ interactions. This mechanism is 
effectively described by introducing a shift of the self-energy of the $\Delta$~\cite{Lee:1981st,Lee:1996mv} $\Sigma_{\pi N}(p_\Delta)$ due to $\Delta \rightarrow \pi N$ in free space
\begin{equation}
\Sigma_{\pi N}(p_\Delta)\rightarrow \Sigma_{\pi N}(p_\Delta)+U_\Delta(\mathbf{p}_\Delta, \rho)
\label{eq:dwidth}
\end{equation}
where $\mathbf{p}_\Delta$ is the three-momentum of the $\Delta$ and $\rho$ is the nuclear density. We generate $U_\Delta(p_\Delta, \rho)$ from a Bruckner-Hartree-Fock calculation using 
a coupled-channel $NN\oplus N\Delta\oplus \pi NN$ model\cite{Lee:1983xu,Lee:1984us,Lee:1985jq,Lee:1987hd}. 
Its real and imaginary part, displayed in Fig~\ref{fig:ud}, exhibit a relatively strong momentum dependence. The decay width of Eq.~\eqref{eq:decay_width} is modified by the imaginary part of $U_\Delta(p_\Delta, \rho)$ as
\begin{align}
\Gamma_\Delta(p_\Delta) \to \Gamma_\Delta(p_\Delta) - 2 {\rm Im}\big[ U_\Delta(p_\Delta, \rho=\rho_0)\big]
\end{align}
where we fixed the density at the nuclear saturation value $\rho_0=0.16$ fm$^3$.

\subsection{Pion-production mechanisms}
\label{pion:prod:sec}
The primary goal of this work consists in further generalizing the factorization ansatz of Eq.~\eqref{eq:factorization} to accommodate productions of real pions in the final state. To this aim, the final state of the reaction can be cast into the form
\begin{align}  
|\psi_f^A \rangle \rightarrow | p_\pi p\rangle \otimes
 |\psi_f^{A-1}\rangle\ , 
\label{eq:factorization_p} 
\end{align}
where $p_\pi$ denotes both the four-momentum $(p^0_\pi,\mathbf{p}_\pi)$ and the isospin $t_\pi$ of the emitted pion. Following the same steps that led to Eq.~(\ref{had:tens}), the incoherent contribution to the one-body one-pion (1b1$\pi$) hadron tensor reads
\begin{align}
W^{\mu\nu}_{\rm 1b 1\pi}({\mathbf q},\omega)=& \int \frac{d^3k}{(2\pi)^3} dE P_h({\mathbf k},E) \frac{d^3p_\pi}{(2\pi)^3} \frac{m_N^2}{e({\mathbf k})e({\mathbf k}+{\mathbf q}-{\mathbf p}_\pi)}\nonumber\\
&\times  \sum_{i}\, \langle k | {j_{i}^\mu}^\dagger |p_\pi p\rangle \langle p_\pi p |  j_{i}^\nu | k \rangle \Big|_{{\mathbf p}={\mathbf k}+{\mathbf q}-{\mathbf p}_\pi}\nonumber\\
& \times \delta(\omega+E-e({\mathbf k}+{\mathbf q}-{\mathbf p}_\pi) + e_\pi(\mathbf{p_\pi}))\,,
\label{had:tens_pi}
\end{align}
where $e_\pi(\mathbf{p_\pi})=\sqrt{{\mathbf p}^2 + m_\pi^2}$ is the
energy of the outgoing pion. Besides the additional integration over
${\mathbf p}_\pi$ the main difference between the above expression and
Eq.~\eqref{had:tens} resides in the elementary amplitude. To describe the
pion-production processes, we need matrix elements of the
charged-current operator causing the transition from a bound
nucleon $|k \rangle$ to a state with a pion and a nucleon
$|p_\pi p\rangle$.

\begin{figure}[h]
\centering
\includegraphics[width=\columnwidth]{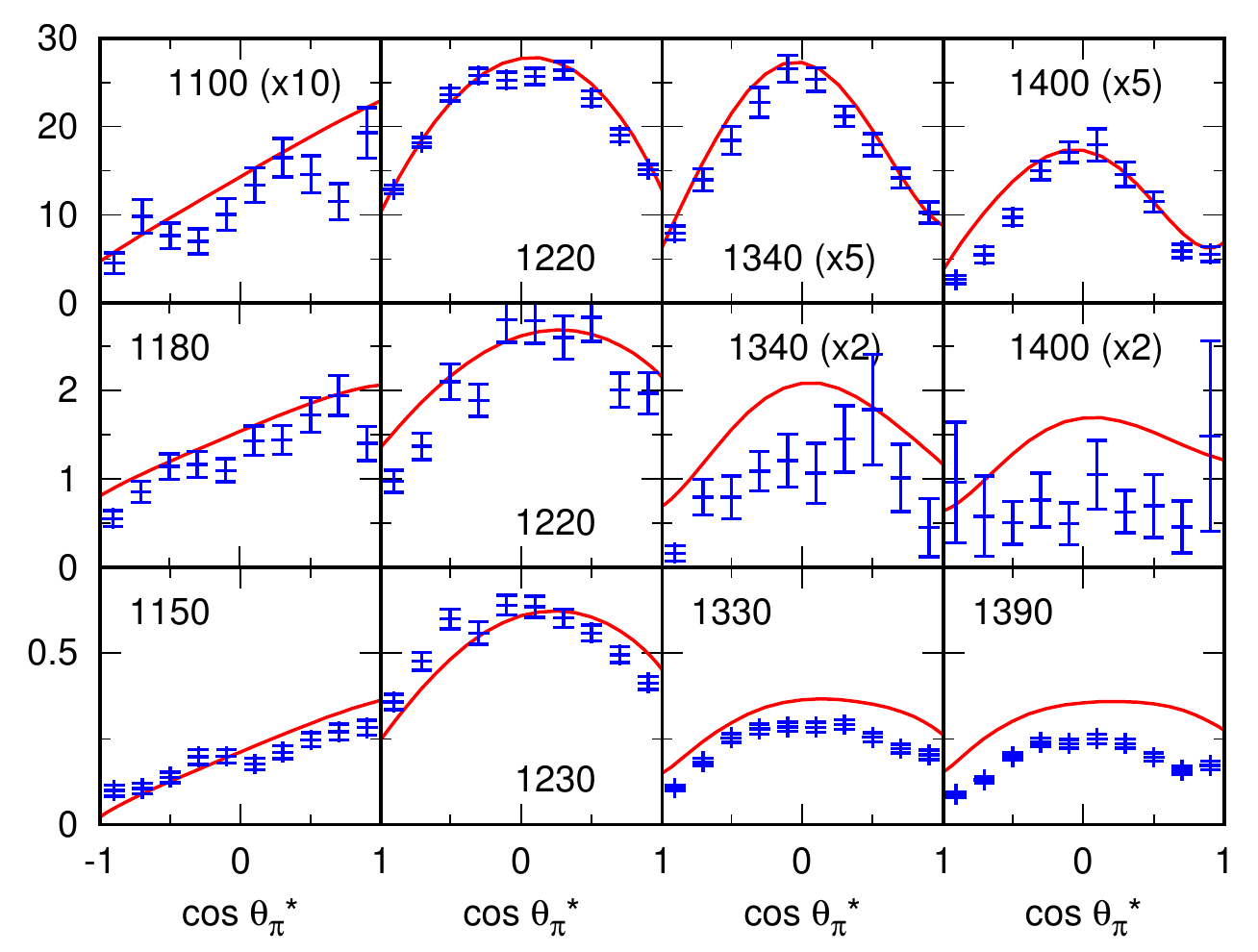}
 \caption{Virtual-photon cross section
 $d\sigma_T/d\Omega^*+\epsilon d\sigma_L/d\Omega^*$ ($\mu$b/sr)
calculated with the DCC model;
$p(e,e'\pi^0)p$.
The top, middle, and bottom rows present the cross sections 
 at $Q^2=0.4$ (GeV/$c$)$^2$, $Q^2=1.76$ (GeV/$c$)$^2$, and
 $Q^2=2.95$ (GeV/$c$)$^2$, respectively.
 In each panel, 
 the number indicates the invariant mass $W$ (MeV),
 and the cross sections are scaled by the factor in the parenthesis.
 Experimental data are from Refs.~\cite{Joo:2001tw,Ungaro:2006df}.}
\label{fig:q0p4}
\end{figure}

\begin{figure}[h]
\centering
\includegraphics[width=\columnwidth]{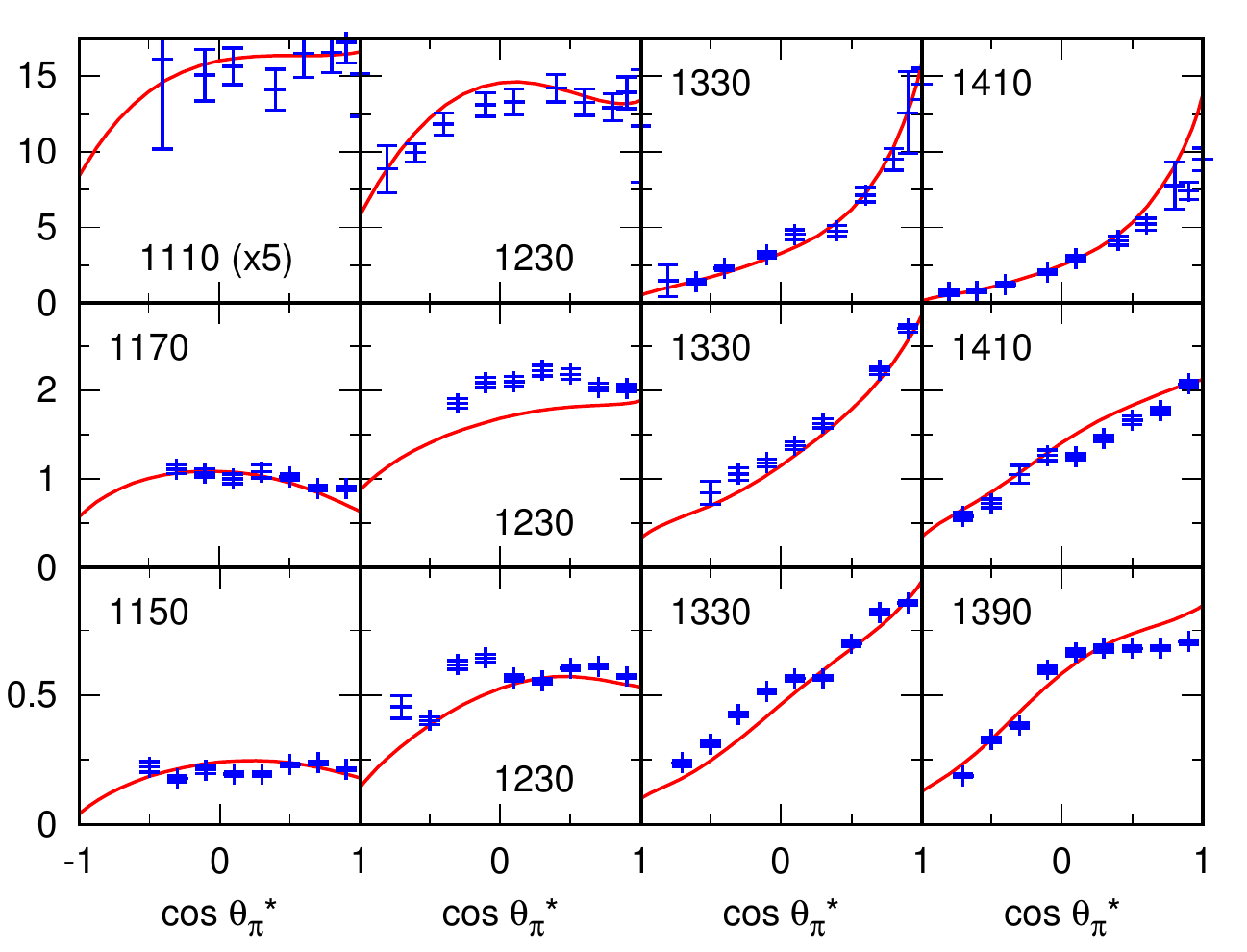}
 \caption{Same as Fig.~\ref{fig:q0p4} but for the $p(e,e'\pi^+)n$ reaction.
Experimental data are from Refs.~\cite{Egiyan:2006ks,Park:2007tn,Park:2014yea}.}
\label{fig:q0p4_2}
\end{figure}

In this work, we employ the ANL-Osaka  coupled-channel 
model~\cite{Kamano:2013iva,Nakamura:2015rta,Kamano:2016bgm}
to generate the current matrix elements $\langle p_\pi p |  j_{i}^\nu | k \rangle$ of Eq.~(\ref{had:tens_pi}).
The ANL-Osaka model is defined by a  Hamiltonian of the following form:
\begin{eqnarray}
H_{\rm AO}= H_0+ \sum_{c,c'}v_{c,c'} +
\sum_{N^*}\sum_{c}[\Gamma_{N^*,c}+\Gamma^\dagger_{N^*,c}] \ ,
\label{eq:ao-h}
\end{eqnarray}
where $H_0$ is the free Hamiltonian, $\Gamma_{N^*,c}$ is a vertex defining
the formation of a bare $N^*$ state
from a meson-baryon channel $c$. The channels included are
$c,c'=\gamma N, \pi N, \eta N, K\Lambda, K\Sigma$, and $\pi\pi N$ with resonant
$\pi\Delta,\rho N$, and $ \sigma N$ components. The energy independent meson-exchange 
potentials $v_{c,c'}$ are derived from 
phenomenological Lagrangians by using  the unitary transformation method~\cite{Sato:1996gk,Kobayashi:1997fm}.
The parameters of the Hamiltonian $H_{\rm AO}$ have been determined in Refs.~\cite{Kamano:2013iva,Kamano:2016bgm}
by fitting about 26,000 data points
of the $\pi N, \gamma N\to \pi N, \eta N, K\Lambda, K\Sigma$ data
from the channel thresholds to $W\le 2.1$~GeV.
The resulting model generates about 20 nucleon resonances which include all of 
 the 4-stars resonances listed by the 
Particle Data Group~\cite{PDG}. 
Here we note that the Hamiltonian in Eq.~(\ref{eq:ao-h}) is consistent with
the conventional nuclear Hamiltonian given in Eq.~(\ref{NMBT:ham}). Thus it can be used straightforwardly to  
generate the current matrix elements $\langle p_\pi p |  j_{i}^\nu | k \rangle$ of Eq.~(\ref{had:tens_pi}).

\begin{figure}[h]
\centering
\includegraphics[width=\columnwidth]{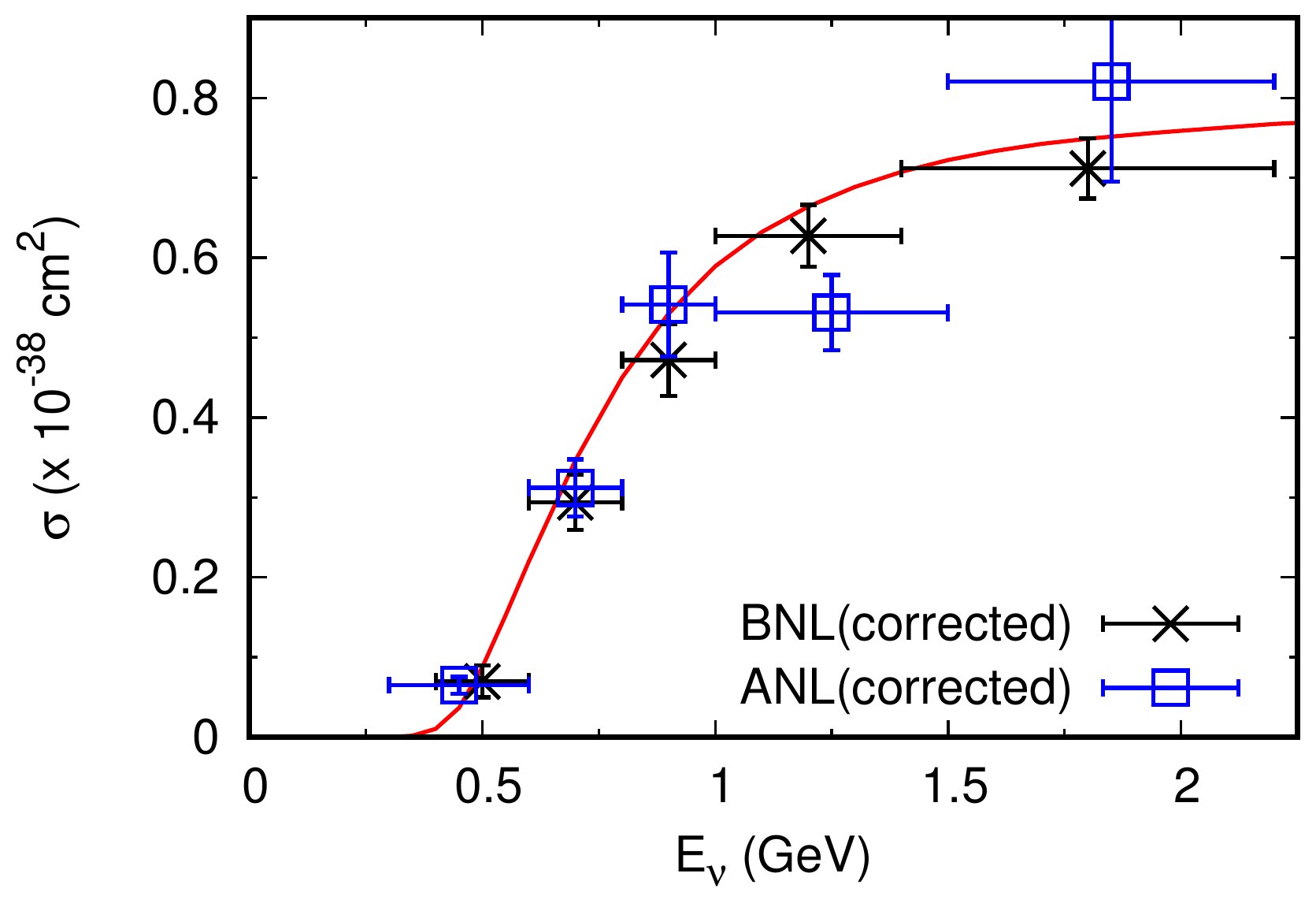}
\includegraphics[width=\columnwidth]{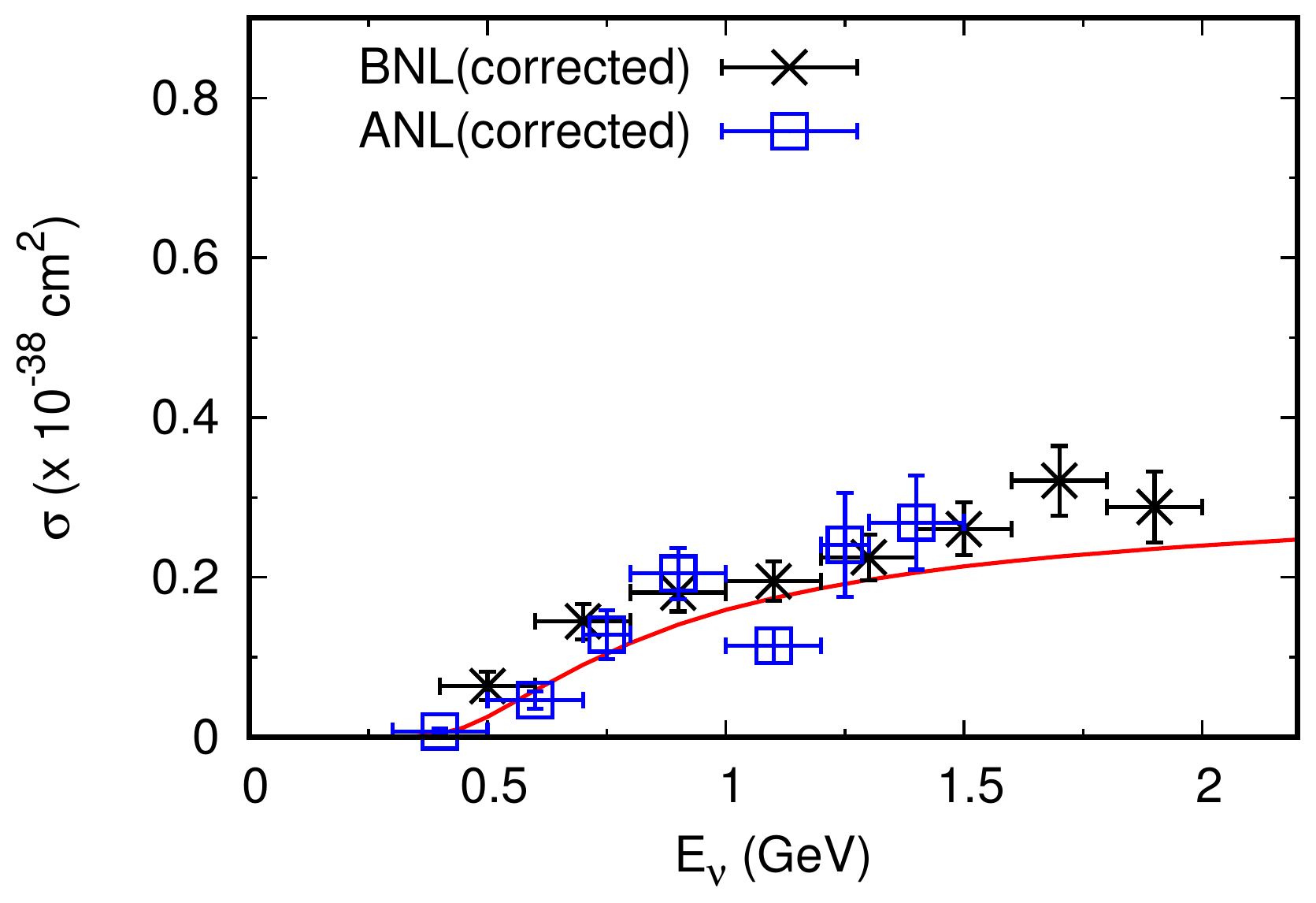}
\includegraphics[width=\columnwidth]{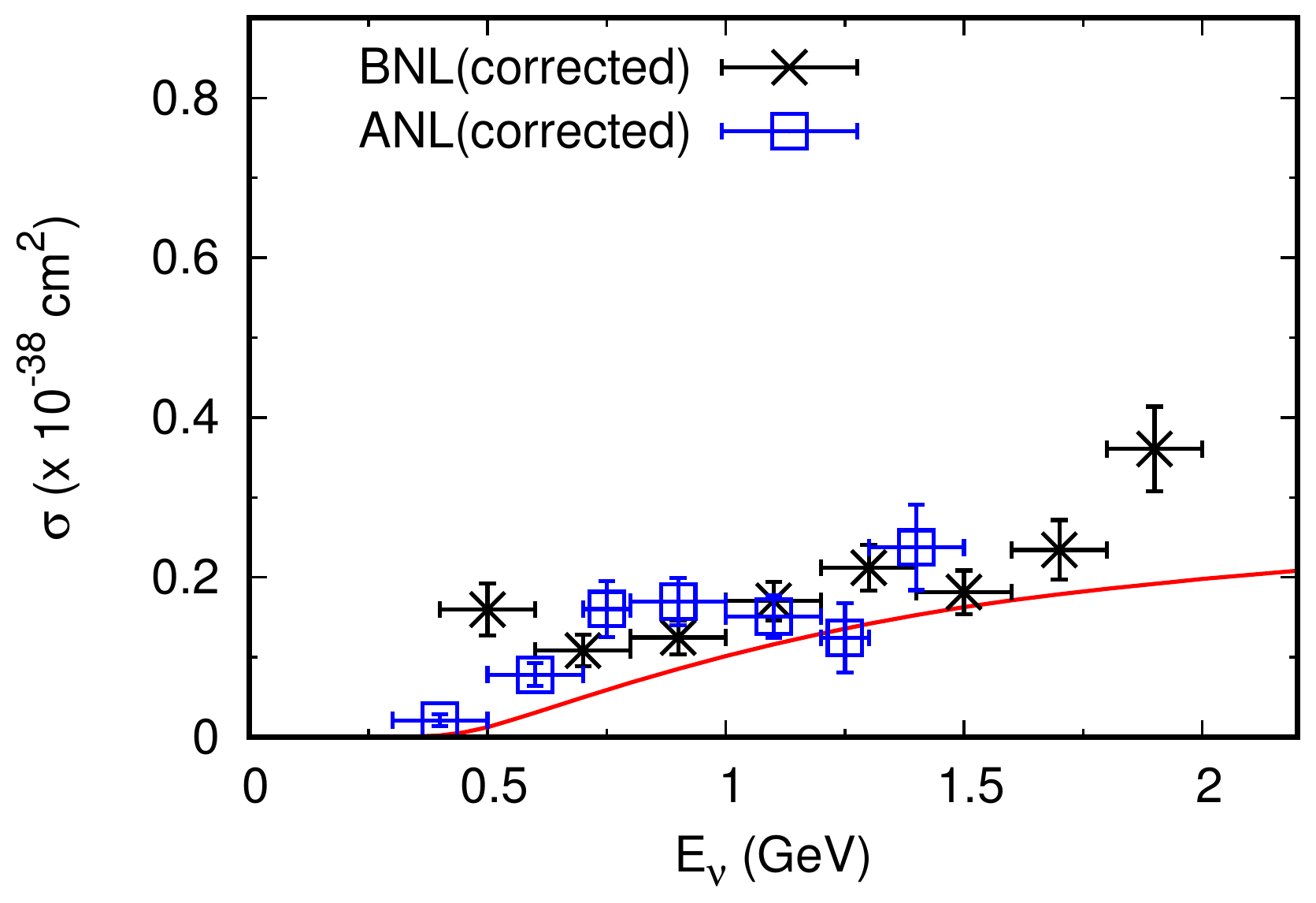}
 \caption{Total cross sections of
(a) $\nu_\mu p \rightarrow \mu^-\pi^+p$;
(b) $\nu_\mu n \rightarrow \mu^-\pi^0p$;
(c) $\nu_\mu n \rightarrow \mu^-\pi^+n$.
The solid red curves are from the DCC model. 
The data are from Ref.~\cite{Rodrigues:2016xjj} where the
ANL~\cite{Barish:1978pj} and BNL~\cite{Kitagaki:1986ct} data have been corrected for the flux uncertainty.
 }
\label{fig:nu-tot}
\end{figure}
\begin{figure}[H]
\centering
\includegraphics[width=\columnwidth]{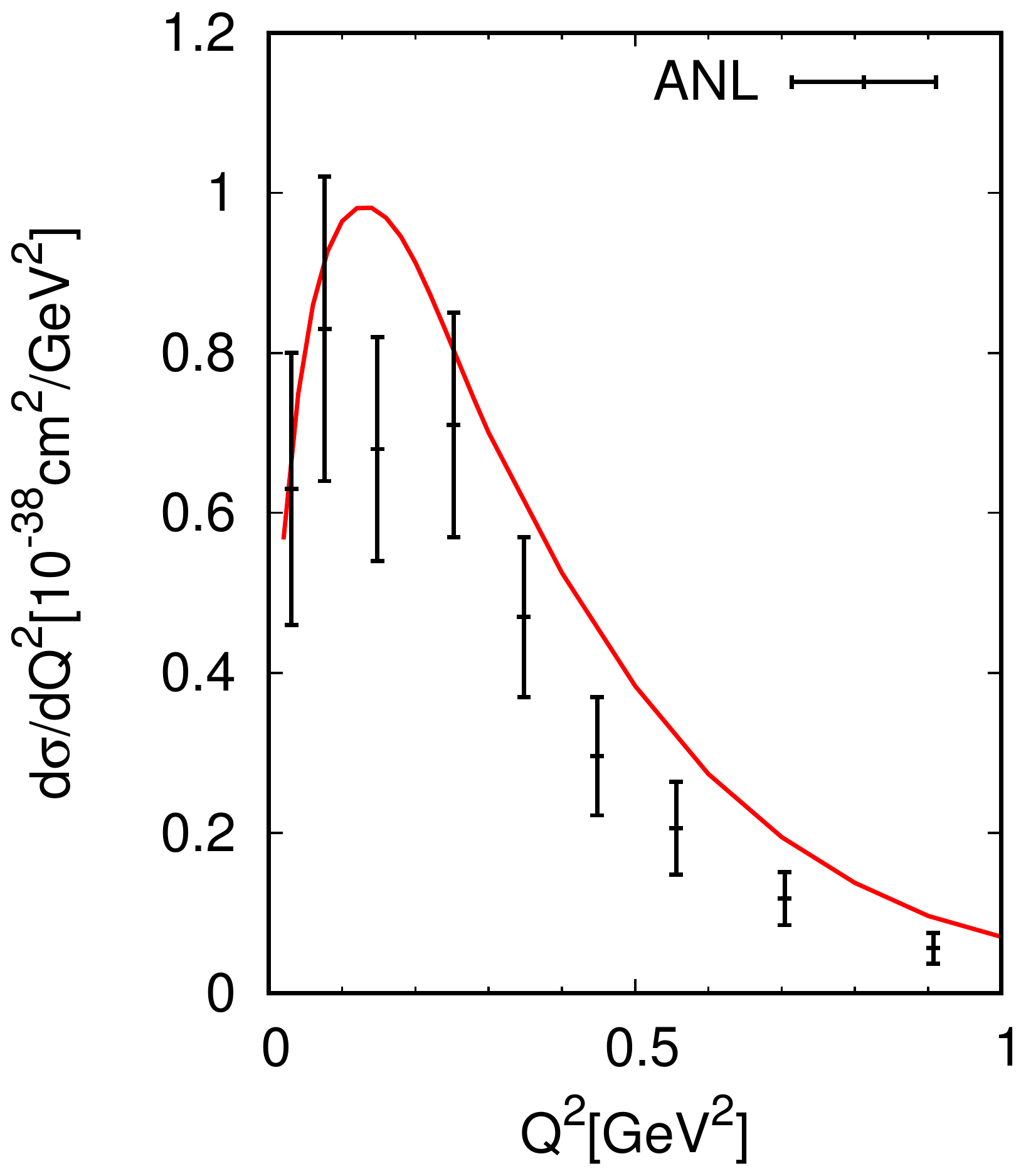}
\includegraphics[width=\columnwidth]{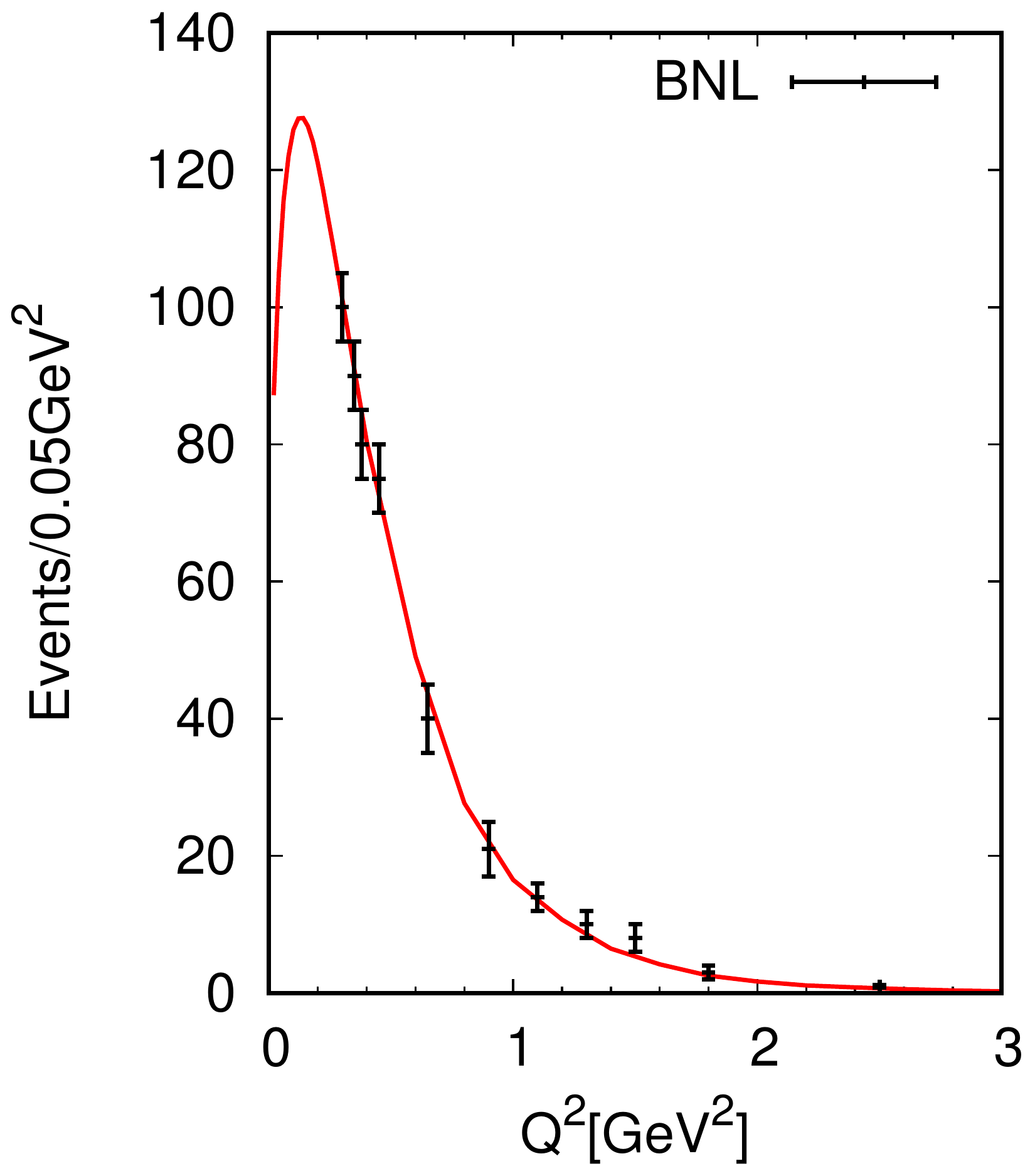}
\caption{Flux-averaged (0.5~GeV $\leq E_\nu \leq$ 6~GeV) $d\sigma/dQ^2$
 for $\nu_\mu p \rightarrow \mu^-\pi^+p$.
 The solid red curves are from the DCC model. 
The data are from
ANL~\cite{Barish:1978pj} and BNL~\cite{Kitagaki:1986ct}.
}
\label{fig:q2-dep}
\end{figure}

The ANL-Osaka model was then extended to the electron- and neutrino-induced
reactions~\cite{Nakamura:2015rta,Nakamura:2016cnn}.
The $Q^2$-dependence of the vector current has been determined by
analyzing data for single-pion electroproduction and inclusive electron
scattering. As an example, we show in Fig.~\ref{fig:q0p4} and Fig.~\ref{fig:q0p4_2} that the ANL-Osaka model
can reasonably describe the data of $p(e,e'\pi^0)p$ and $p(e,e'\pi^+)n$ reactions, respectively,
for $Q^2= 0.40$ (GeV/c)$^2$(top),  $Q^2=1.76$ (GeV/c)$^2$ (middle),
 $Q^2=2.95$ (GeV/c)$^2$ (bottom).

On the other hand, the axial current associated with nucleon resonances cannot be determined
very well because the neutrino-induced meson production data are scarce
except in the $\Delta(1232)$-region.
Thus we determined the axial couplings using the PCAC relation to the
$\pi N$ reaction amplitudes, and assumed the dipole $Q^2$-dependence
with the cutoff of $\simeq 1$ GeV. In Figs.~\ref{fig:nu-tot} and
\ref{fig:q2-dep}, we show that the neutrino data for the total cross sections and
the $Q^2$-dependence of
the single pion production can be described very well by the ANL-Osaka model.

Here we note that
the  PCAC relation with the $\pi N$ amplitudes, and in particular with their
phases, is not taken into account in other pion production models,
such as the Rein-Sehgal model~\cite{Rein:1980wg} -- commonly used in analyzing
neutrino experiments -- and the LPP model~\cite{Paschos:2003qr,Lalakulich:2005cs,Lalakulich:2006sw}
recently employed to calculate inclusive processes~\cite{Vagnoni:2017hll} within the 
CBF hole SF formalism. 
This inconsistency leads to significant differences in the structure
function $F_2$ at $Q^2\sim 0$~\cite{Nakamura:2015rta}.

Prior to the present work,
the ANL-Osaka DCC model has been applied to electroweak reactions on the simplest
nucleus, the deuteron~\cite{Nakamura:2017sls,Nakamura:2018fad,Nakamura:2018ntd,Nakamura:2018cst}.
Predictions from 
the DCC-based model, which includes the impulse as well as $NN$ and
meson-nucleon rescattering mechanisms, agree reasonably well with the
data on $\gamma d\to \pi NN$~\cite{Nakamura:2018fad,Nakamura:2018cst}
and $\gamma d\to \eta NN$~\cite{Nakamura:2017sls}.

The model was also used to study final state interaction (FSI) effects
on $\nu_\mu d\to \mu^- \pi NN$~\cite{Nakamura:2018cst}, leading to
the FSI corrections to the 
ANL~\cite{Barish:1978pj} and BNL~\cite{Kitagaki:1986ct} data for
 $\nu_\mu N\to \mu^- \pi N$ which had been extracted from the deuteron
target data without correcting for the significant FSI effects.

Analogously to these studies on the deuteron, in this work the DCC 
amplitudes in the laboratory frame are obtained by boosting the corresponding
ones in the center-of-mass frame, where the DCC model was originally developed.
Here we briefly describe  the procedures for  calculating the  current  matrix elements $<p_\pi p|j^\nu_i(q)|k>$ 
in Eq.~(\ref{had:tens_pi})   from those evaluated in the center of mass (CM) frame of $\pi N$. 
Including explicitly
the nucleon spin quantum numbers $m_s$, we can write
\begin{eqnarray}
&& <p\,\,m_{s'}, p_\pi|j^\nu_i(q)|k\,\, m_s>
\nonumber \\
&=&
\sqrt{\frac{E_\pi({\mathbf k_c})E_N(-{\mathbf k_c})}
     {E_\pi({\mathbf p}_\pi)E_N({\mathbf p})}}
\sqrt{\frac{|{\mathbf q_c}|E_N({-\mathbf q_c})}{|{\mathbf q}|E_N({\mathbf k})}}
\sum_{\mu}\Lambda_\mu^\nu(p_t)
\nonumber\\
&&\times
 \Big[\sum_{m'_{s_c},m_{s_c}}
 R^*_{m_{s'_c},m_{s'}}(p,p_t)
R_{m_{s_c},m_s}(k,p_t) \nonumber \\
&&\times \langle\pi({\mathbf k_c}), N(-{\mathbf k_c}\, m'_{s_c})|j^\mu_{i}(q_c)| N(-{\mathbf q_c} \, m_{s_c})\rangle\Big]
\label{eq:amp_imp_lor-a}
\end{eqnarray}
where the suffixes '$ c$' indicate quantities in the CM system of $\pi N$,
$p_t =p+ p_\pi = q + k$ is the total four-momentum of the $\pi N$ system, defined by
${\mathbf p_t} ={\mathbf p}+{\mathbf p_\pi}={\mathbf q}+{\mathbf k}$ and
$p_t^0=E_{N}({\mathbf p})+E_\pi({\mathbf p_\pi})=\omega+E_N({\mathbf k})$.
The CM matrix elements of the current operator, 
$\langle \pi({\mathbf k_c}), N(-{\mathbf k_c}\, m'_{s_c})|j^\mu_{i}(q_c)| N(-{\mathbf q_c} \, m_{s_c})\rangle$
, from the ANL-Osaka model 
are calculated following the procedure 
detailed in Appendix~D of Ref.~\cite{Kamano:2013iva}.

In Eq.~(\ref{eq:amp_imp_lor-a}), the quantity $\Lambda^\nu_\mu(p_t)$ boosts 
any  momentum $a_{c}=(a^0_c, {\bm a_c})$ in the CM of the considered $\pi N$ system
 to the momentum  
$a_L=(a^0_L, {\bm a_L})$ in the laboratory frame by the following  Lorentz transformation :
\begin{eqnarray}
a_{L}^0 &=&\sum_{\nu}\Lambda^0_\nu(p_t) a^\nu_{ c}
= \frac{a_{ c}^0\, p_t^0 + {\mathbf p}_t\cdot{\bm a}_c} { m_t}\ ,
\nonumber \\
a_{L}^i &=& \sum_{\nu}\Lambda^i_\nu(p_t) a^\nu_{ c}
=a_{ c}^i + p_t^i\left[\frac{ {\mathbf p}_t\cdot {\bm a}_{ c}}{ m_t (m_t+p_t^0)}
+\frac{ {a}^0_{c}}{ m_t}\right] \ ,
\label{eq:lorentz}
\end{eqnarray}
where the index $i=1,2,3$ is a spatial component and $m_t\equiv\sqrt{p_t \cdot p_t}$.

The spin rotation matrix $R_{s_{\bar c},s}(p,p_t)$ in Eq.~(\ref{eq:amp_imp_lor-a}) 
is given~\cite{Polyzou:2014yea,Polyzou:2014dja,Polyzou:2014priv}
explicitly as:
\begin{eqnarray}
&&R_{m_{s_c},m_s}(p,p_t) = \nonumber\\
&& \langle m_{s_c}| B^{-1}(p_{c}/m_N)B^{-1}(p_t/m_t)B(p/m_N) |m_s \rangle \ ,
\label{eq:spin-rot}
\end{eqnarray}
where $|m_s\rangle$ is the nucleon spin state,
$p_{ c}$ is obtained from the nucleon momentum $p$ in the laboratory frame by
the Lorentz transformation of Eq.~(\ref{eq:lorentz}),
and
\begin{eqnarray}
B (p/m) &=& \frac{1}{\sqrt{2m\,(p^0+m)}} ((p^0+m){\mathbf I} +
{\mathbf p}\cdot {\mathbf \sigma}) \ , \nonumber \\
B ^{-1}(p/m) &=& \frac{1}{\sqrt{2m\,(p^0+m)}} ((p^0+m){\mathbf I} -
{\mathbf p}\cdot {\mathbf \sigma}) \ ,
\label{eq:lb-tx}
\end{eqnarray}
where ${\mathbf \sigma}$ is the Pauli operator and ${\mathbf I}$ is the unit matrix.

\section{Results}
\label{sec:res}
In the same manner as Ref.~\cite{Rocco:2018mwt}, the numerical integration of Eqs.~(\ref{had:tens}), (\ref{had:tens2}), (\ref{had:tens_pi}) are carried out by means of a dedicated Metropolis Monte Carlo algorithm. Since the integrands extend up to large momentum and removal energy, when evaluating $W^{\mu\nu}_{\rm 1b}$ and $W^{\mu\nu}_{\rm 1b 1\pi}$ it is convenient to employ a normalized hole-SF as the importance-sampling function.  Analogously, the importance-sampling function of choice for $W^{\mu\nu}_{\rm 2b}$ is proportional to the product $P_h^{\rm NM}({\bf k},\tilde{E})P_h^{\rm NM}({\bf k}^\prime,\tilde{E}^\prime)$.

\begin{figure}[t]
\centering
\includegraphics[width=\columnwidth]{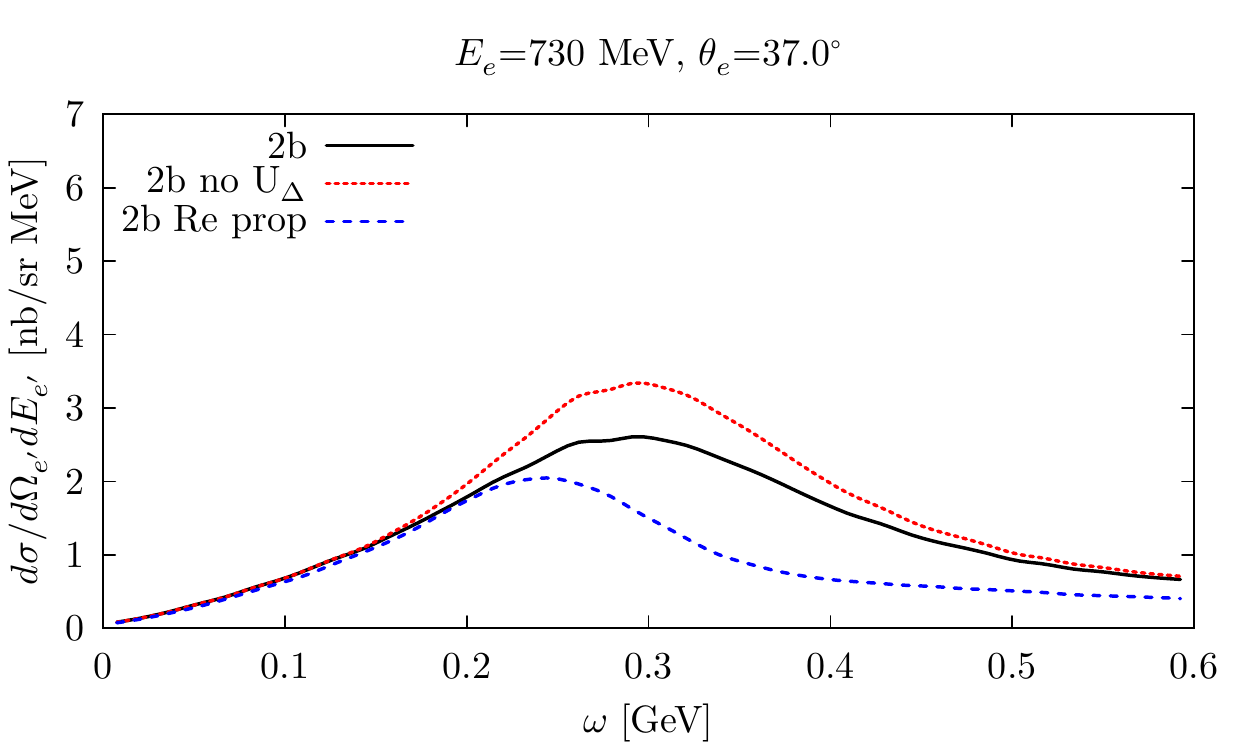}
\caption{Two-body current contribution to the double-differential electron-$^{12}$C cross section for $E_e=730$ MeV and $\theta_e= 37^\circ$. 
The solid (black) line corresponds to results in which the in-medium
 corrections to the $\Delta$-decay are included, while the short-dashed
 (red) line is obtained neglecting this contribution. The dashed (blue) line 
 displays the two-body current contribution in which only the real part of the
 $\Delta$  propagator is retained.}
\label{fig:twobd}
\end{figure}

\begin{figure}[h]
\centering
\includegraphics[width=\columnwidth]{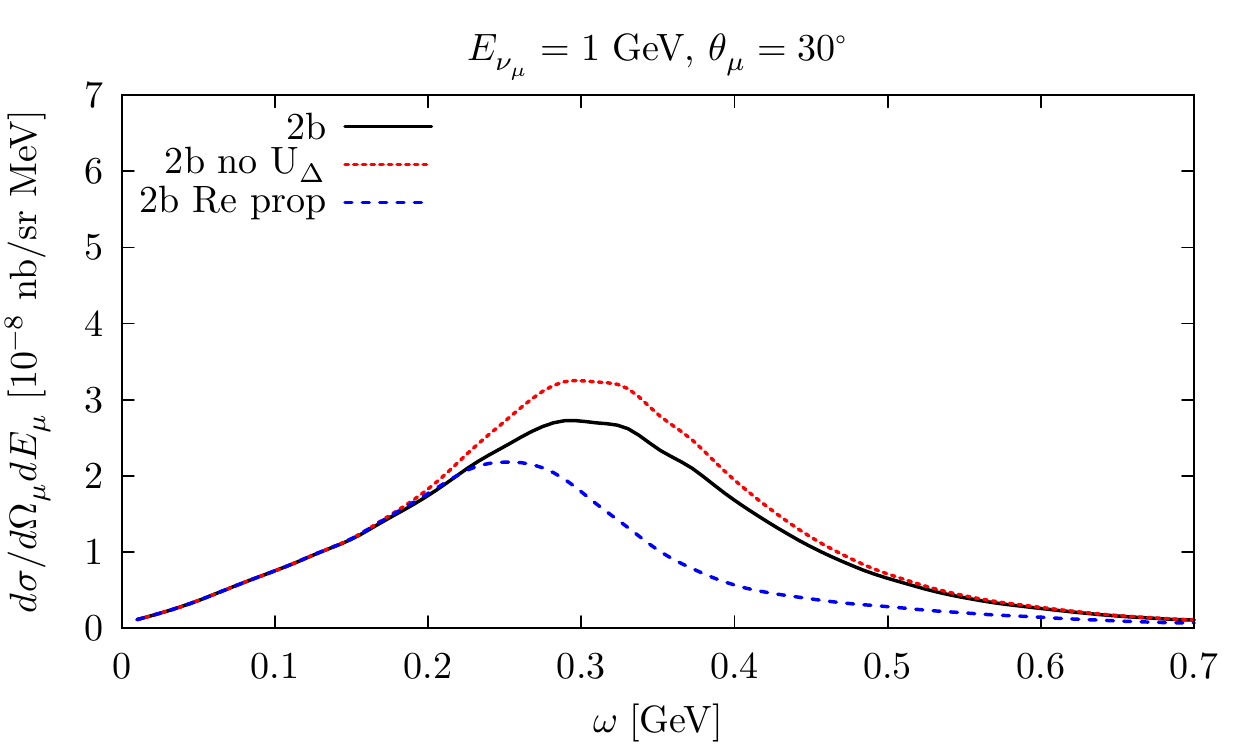}
\caption{Same as Fig.~\ref{fig:twobd} but for CC $\nu_\mu$-$^{12}$C scattering at $E_{\nu_\mu}=1$ GeV and $\theta_\mu=30^\circ$.}
\label{fig:meffect}
\end{figure}

Fig.~\ref{fig:twobd} shows the MEC contribution to the double-differential electron-$^{12}$C cross section for $E_e=730$ MeV and $\theta_e=37^\circ$. The solid (black) line corresponds to the full calculation in which the in medium $\Delta$-potential $U_\Delta$ has been included in the propagator, as explained in Sec.~\ref{mec:sec}. On the other hand, the short-dashed (red) line is obtained disregarding this contribution. The comparison between the two curves clearly shows that accounting for the in-medium decay of the $\Delta$ leads to a visible quenching of the MEC contribution to the inclusive cross section. A similar behavior is also observed in weak processes, as shown in Fig.~\ref{fig:meffect}, where the effects of the in-medium potential of the $\Delta$ is analyzed in the CC $\nu_\mu$-$^{12}$C scattering cross section for a beam energy $E_\nu=1$ GeV and scattering angle $\theta_\mu=30^\circ$. In this particular kinematical setup, including $U_\Delta$ brings about a $\simeq 15\%$ depletion of the MEC strength. The way we include medium effects on the $\Delta$ propagation is significantly different from the prescription of keeping only the real part of $\Delta$ propagator~\cite{DePace:2003spn,Simo:2016ikv,Butkevich:2017mnc}, leading to the dashed (blue) lines of Figs.~\ref{fig:twobd} and \ref{fig:meffect}. Disregarding altogether the imaginary-part of the $\Delta$ propagator brings about a stronger reduction of the strength than including $U_\Delta$. In addition, the position of the peak is shifted to lower energy transfers.

\begin{figure}[b]
\centering
\includegraphics[width=\columnwidth]{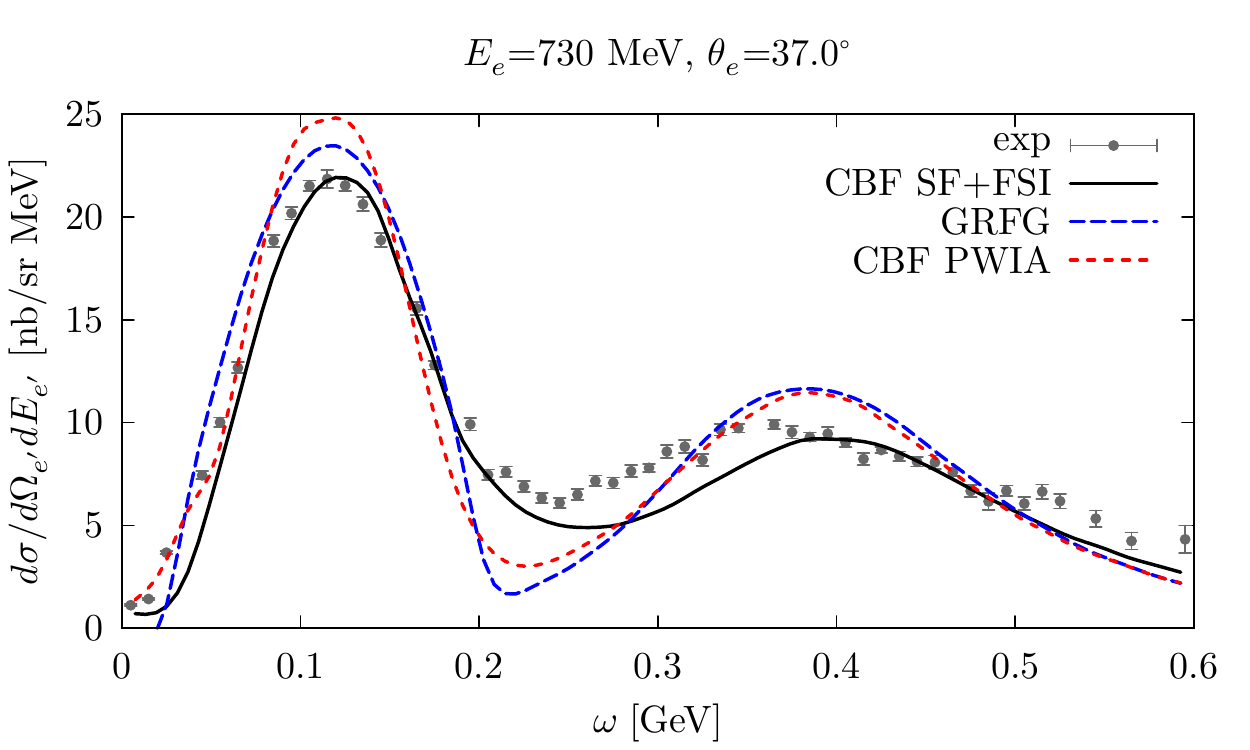}
\caption{Electron-$^{12}$C double-differential cross section. The dashed (blue) line has been obtained within the GRFG model.
The short-dashed (red) and solid (black) curves have been obtained using the SF of Ref.~\cite{Benhar:1994hw} within the PWIA and IA with FSI corrections, respectively.} 
\label{fig:compare}
\end{figure}

In Fig.~\ref{fig:compare} we compare the results obtained for the
electron-$^{12}$C scattering double differential cross section for
$E_e=730$ MeV and $\theta_e= 37^\circ$ employing different
approximations to describe the nuclear target and the final state
interactions. It has to be noted that, when computing the MEC
contribution, the two-body hole SF is approximated by the product of two
one-body hole SF, as in Eq.~\eqref{had:tens2}. The cross sections with a
real pion in the final state are computed convoluting the DCC elementary
amplitudes with the one-nucleon SF, as discussed in
Sec~\ref{pion:prod:sec}, and a cut on invariant energies $W \leq
2.0$~GeV has been applied.

The dashed (blue) curve has been obtained using the global relativistic Fermi gas (GRFG)
model, which only entails statistical correlations, to determine the hole SF
\begin{equation}
P_h^{\rm GRFG}({\bf k},E)= \theta(k_F-|\mathbf{k}|) \delta\left(E+\frac{k^2}{2m}\right)\, .
\end{equation}
As for the Fermi-momentum, we take $k_F$= 225 MeV and no binding energy is introduced. The short-dashed (red) line displays the Plane Wave Impulse Approximation (PWIA) result in which the excitation energies of the $(A-1)$-body spectator system are assumed to be constant, $E_f^{A-1}=\bar{E}^{A-1}$. Hence, the hole SF reduces to
\begin{equation}
P_h^{\rm PWIA}({\bf k},E)= n_h({\bf k}) \delta(E+\bar{E}^{A-1}-E_0^{A})\, ,
\end{equation}
thereby loosing information on the removal-energy distribution of the
target. The momentum distribution employed in the PWIA calculations,
represented by the black solid line of Fig.~\ref{fig:nk}, is derived by
integrating over the removal energy of the CBF hole SF of Ref.~\cite{Benhar:1994hw}.

\begin{figure*}[t]
\centering
\includegraphics[width=\columnwidth]{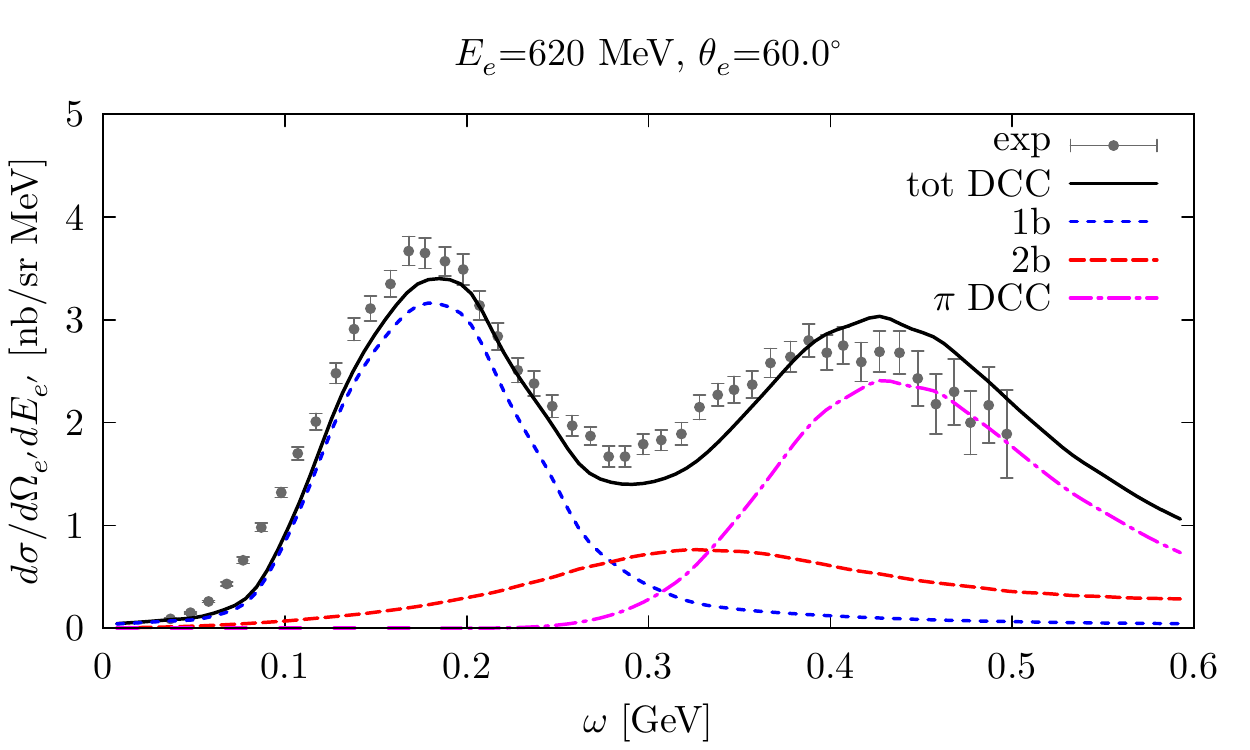}
\includegraphics[width=\columnwidth]{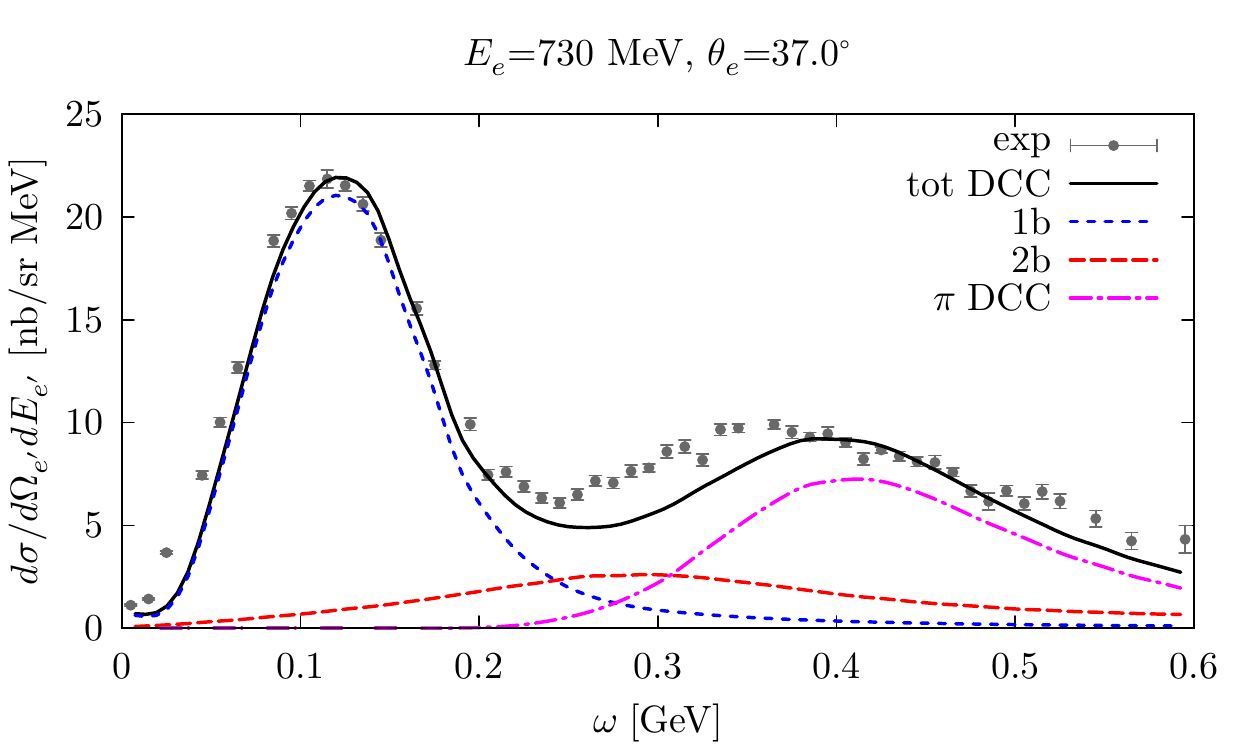}
\includegraphics[width=\columnwidth]{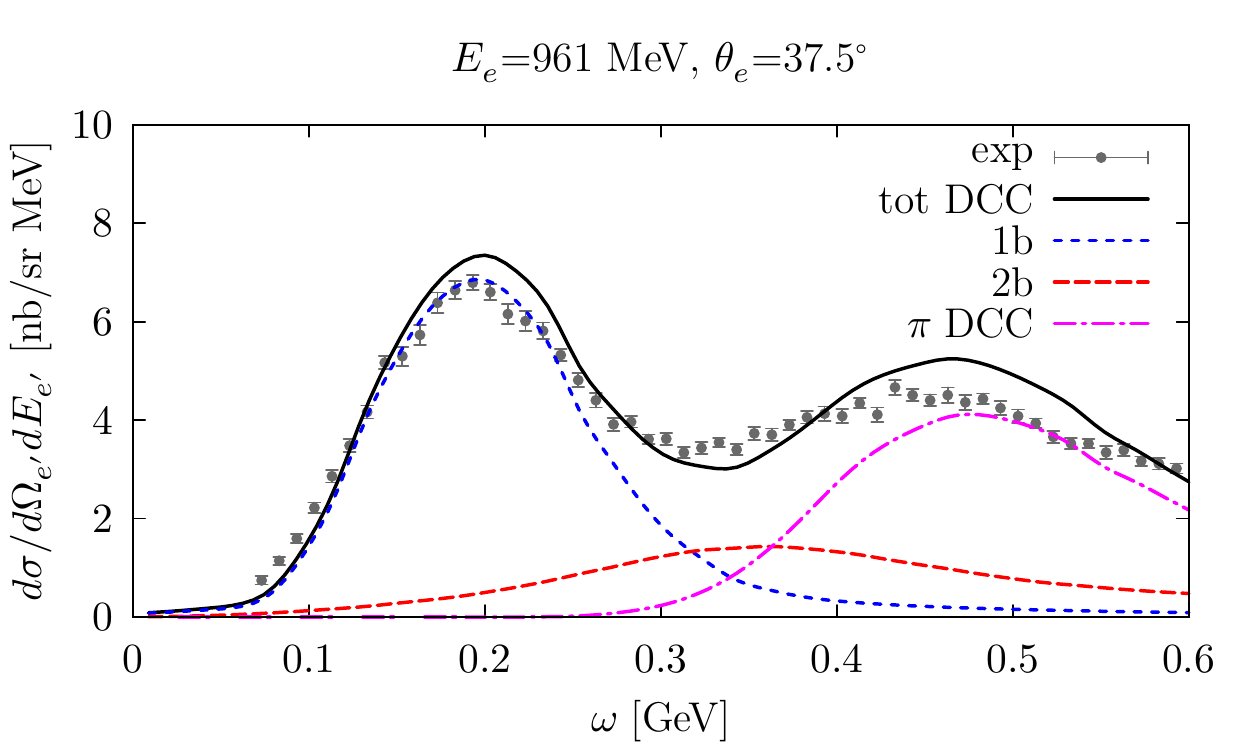}
\includegraphics[width=\columnwidth]{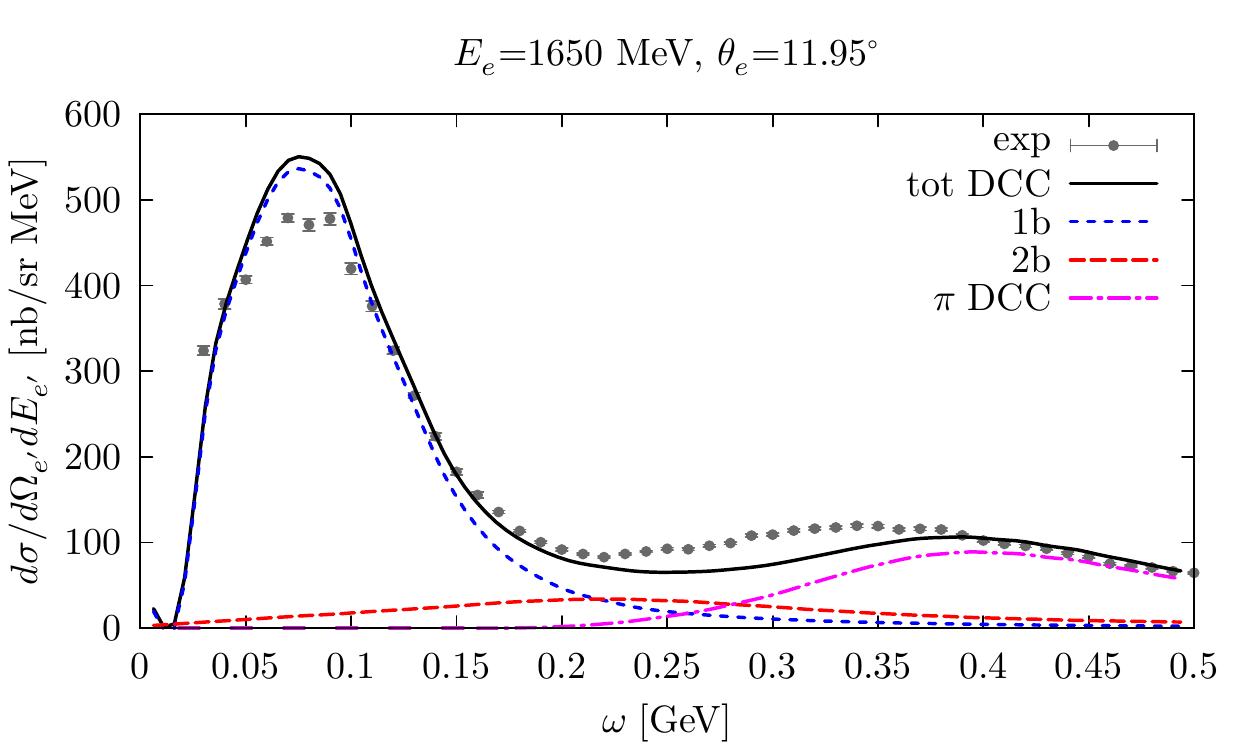}
\caption{ Electron- $^{12}$C inclusive cross sections for different combinations of $E_e$ and $\theta_e$. The short-dashed (blue) line and dashed (red) line correspond to one- and two-body current contributions, respectively.  The dash-dotted (magenta) lines represent $\pi$ production contributions. The solid (black) line is the total results obtained summing the three different terms.}
\label{ee:12C:rho_mec}
\end{figure*} 

The solid (black) line in Fig.~\ref{fig:compare} is obtained using the full CBF hole SF to
describe the quasi-elastic peak and the $\pi$-production regions. For
this most sophisticated treatment of the target nucleus, we also show
results in which the impulse approximation is corrected by including
FSI. In single-nucleon knockout processes, this is achieved following
Eqs.~(\ref{fold:func}--\ref{fold:func:expl}), i.e. employing the real
part of an optical potential derived from the Dirac phenomenological fit
of Ref~\cite{Cooper:1993nx} and the folding function of
Refs.~\cite{Benhar:2006wy,Benhar:2013dq}. The main two consequences of
including FSI are a shift of the the quasielastic peak and a
redistribution of the strength towards lower values of $\omega$. In
two-nucleon emission processes, FSI are effectively accounted by
including in their energy spectrum a momentum-independent binding of $60$ MeV
 per particle.
Treating FSI with the same level of
sophistication as for the one-nucleon knockout requires the knowledge of
the optical potential associated to the removal of two-nucleons from
$^{12}$C and the corresponding folding functions. In addition,
single-charge exchange processes~\cite{Colle:2015lyl} and interactions
taking place within the pair of struck nucleons should also be properly
modeled. FSI between the $\pi$-nucleon state and the $A-1$ spectator
system are not addressed in this article.

For exclusive single pion production processes from neutrino-${}^{12}$C 
scattering in the $\Delta(1232)$ region, it has been shown that pion absorptions and 
redistribution of the pion momentum spectrum
are important FSI effects~\cite{Hernandez:2013jka,Lalakulich:2012cj}.
However, by definition, the (semi-)classical treatments of the FSI therein employed do not 
modify the inclusive observables analyzed in the present work. 
A more systematic treatment of FSI in processes with both two outgoing nucleons and a pion and a nucleon in the final state is currently being investigated and will be the subject of a forthcoming work.

By comparing the solid with the dashed and short-dashed lines it clearly emerges that an accurate treatment of nuclear dynamics in the initial state and the inclusion of FSI considerably improve the agreement with experimental data in the whole energy-transfer region. For this particular kinematical setup, neglecting the correlations between the removal energy and momentum, as in the PWIA, leads to an overshooting of the quasi-elastic peak, even compared to the crudest GRFG model. This is consistent with Ref.~\cite{Sobczyk:2017vdy} where the use of a realistic hole SF was found to produce noticeably different scaling features of the nucleon-density response from those obtained within the simple PWIA.

Figure~\ref{ee:12C:rho_mec} displays the double-differential electron-$^{12}$C cross sections in four kinematical setups, corresponding to: $E_e=620$ MeV, $\theta_e=60^\circ$ (upper-left panel), $E_e=730$ MeV, $\theta_e=37^\circ$ (upper-right panel), $E_e=961$ MeV, $\theta_e=37.5^\circ$ (lower-left panel), and $E_e=1650$ MeV, $\theta_e=11.95^\circ$ (lower-right panel). The total cross section, represented by the solid (black) line, is obtained as in Fig.~\ref{fig:compare} using the CBF hole-SF of Ref.~\cite{Benhar:1994hw} and including FSI as discussed above. The breakdown of the contributions associated with the different reactions mechanisms is also shown. The dashed (blue) line is the quasi-elastic peak obtained including the one-body current only, while the
short-dashed (red) line corresponds to two-nucleon knockout final states induced by MEC reaction mechanisms. The cross section associated with the emission of a real pion and a nucleon is represented by the dot-dashed (magenta) line. 

In all kinematical setups, MEC enhance the cross section primarily in
the dip region, between the quasielastic and the $\Delta$ peaks. Their 
strength exhibits a strong dependence on the electron scattering angle; it increases relatively
to the one of one-body processes for larger values of the scattering
angle. This is consistent with the findings of Ref.~\cite{Rocco:2018mwt} and can be
traced back to the fact that two-body currents are most effective in
transverse responses. Note that, as discussed in
Sec.~\ref{mec:sec}, the interference between one- and two-body currents
is not included in our calculations. Although it was argued in
Ref.~\cite{Benhar:2015ula} that this leads to a small enhancement in the
dip region within the factorization scheme, GFMC calculations have demonstrated 
that the interference contribution significantly increases the transverse 
electroweak responses~\cite{Lovato:2016gkq,Lovato:2017cux}.

There is an overall good agreement between theoretical results and
experimental data in all the kinematical setups we considered. In
particular, the inclusion of realistic pion production mechanism turns out
to be essential to reproduce the data in the $\Delta$-production region.
Comparing our findings with those of Ref.~\cite{Rocco:2015cil},
it appears that the DCC model largely overcomes the limitations of the structure
functions of Ref.~\cite{Bodek:1980ar} in describing the region of  $Q^2 \lesssim 0.2$ GeV$^2$.
The remaining discrepancies between our theoretical calculations and experiments
are most likely due to the in-medium broadening of the $\Delta(1232)$~\cite{Nakamura:2009iq}, which is missing
in the present version of the DCC model.  
The MEC may also need to be refined by, for example, carefully analyzing the $\gamma d\to pn$ reaction, as has been
done in Ref.~\cite{Chen:1988xq}. Finally, the afore-mentioned missing interference between one-and two-body currents, 
together with a full account of FSI in two-nucleon knockout and pion-production processes are all needed to further improve
the agreement with experiment. All these points will be addressed in future work.

\begin{figure}[H]
\centering
\includegraphics[width=\columnwidth]{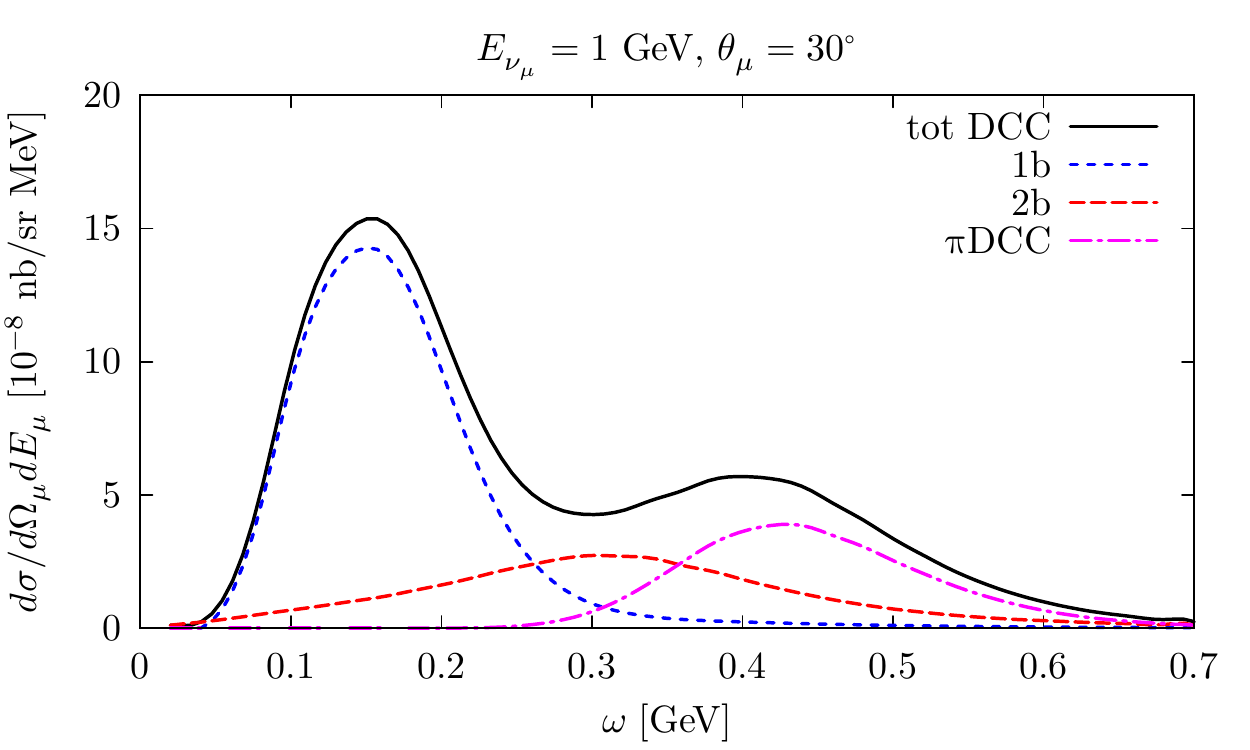}
\includegraphics[width=\columnwidth]{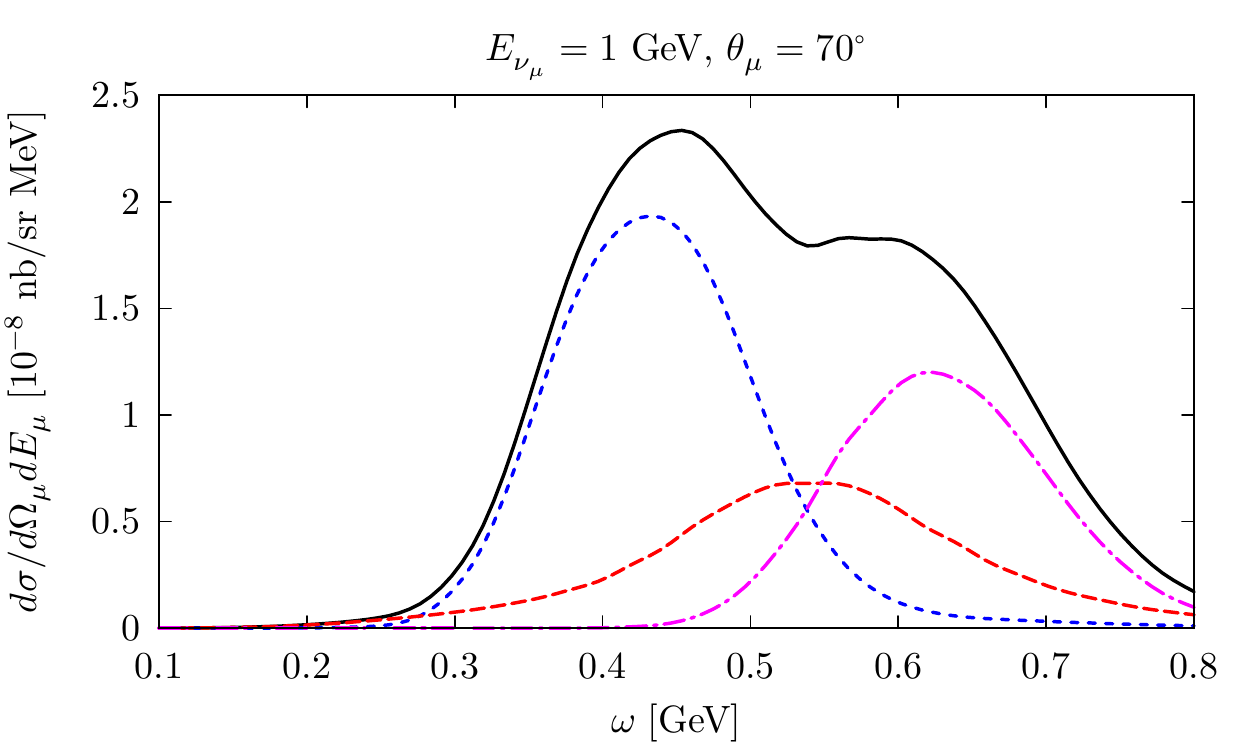}
\caption{ Double-differential cross section for the $\nu_\mu$ + $^{12}$C $\rightarrow$ $\mu^-$ + X process at $E_\nu=1$ GeV, $\theta_\mu=30^\circ$ (upper panel), and $E_\nu=1$ GeV, $\theta_\mu=70^\circ$ (lower panel). The different curves are the same as in Fig.~\ref{ee:12C:rho_mec}. }
\label{nu:12C:rho_mec}
\end{figure}

The results obtained for the double-differential CC $\nu_\mu$-$^{12}$C
scattering cross sections are shown in Fig.~\ref{nu:12C:rho_mec} for
$E_\nu=1$ GeV, $\theta_\mu=30^\circ$ (upper panel), and $E_\nu=1$ GeV,
$\theta_\mu=70^\circ$ (lower panel). The calculations have been carried
out within the same framework employed in the electromagnetic case. The
only additional ingredients are the axial terms in the current
operators and in the $\pi$-production amplitudes.
Consistently with the results of
Fig.~\ref{ee:12C:rho_mec} and with Ref.~\cite{Rocco:2018mwt}, the
relative strength of the MEC contribution increases with the scattering
angle, reflecting the primarily transverse nature of this term even when
axial terms are present.
To the best of our knowledge, precise inclusive neutrino double-differential cross section data covering
the $\Delta(1232)$ region are not available, yet. Comparing our theoretical calculations with such data 
requires a convolution with the neutrino energy spectrum of the experiments.
In this work, primarily aimed at demonstrating the possibility of including relativistic one- and two-body
current together with reliable pion-production amplitudes, we refrain from presenting flux-folded results.
To this aim, a more sophisticated treatment of FSI, for both two-nucleon knockout and pion-production processes
is required.

\section{Conclusions}
\label{sec:concl}
We have carried out rigorous calculations of electron- and neutrino-scattering off $^{12}$C in the broad kinematical region of interest for current and planned neutrino-oscillation experiments. The EFS has allowed us to combine a realistic description of nuclear dynamics in both the initial target state and the spectator system -- achieved by employing a SF computed within the CBF theory~\cite{Benhar:2006wy} -- with a relativistic interaction vertex, suitable to include different reaction mechanisms. The QE and ``dip'' regions are modeled by relativistic one- and two-body currents. In-medium modification of the $\Delta$ propagator is accounted for by a phenomenological potential derived within BHF~\cite{Lee:1983xu,Lee:1984us,Lee:1985jq,Lee:1987hd}. The consequent reduction of the MEC strength is less important than the one resulting from the {\it ad hoc} prescription of disregarding the imaginary part of the $\Delta$ propagator~\cite{DePace:2003spn,Simo:2016ikv,Butkevich:2017mnc}. The elementary amplitudes relevant for pion-production processes are obtained within the ANL-Osaka DCC model~\cite{Kamano:2013iva,Nakamura:2015rta,Kamano:2016bgm}, which contains about 20 nucleon resonances, can be reliably utilized up to an invariant mass of $W\le 2.1$~GeV. Their numerical implementation has required a further development of our highly-parallel Metropolis Monte Carlo integration technique. 

To quantitatively assess the role of realistic hole-SF and FSI effects, we first computed the electron-$^{12}$C double-differential cross sections for incoming energy $E_e = 730$ MeV and scattering angle $\theta_e=37^\circ$. An accurate treatment of nuclear dynamics in both the initial and final states is required to reproduce experimental data. In particular, both the GRFG model and the simplest version of the PWIA -- in which the excitation energies of the spectator system are assumed to be constant -- noticeably overestimate the strength of the quasi-elastic peak. We have carried out calculations for the electron-$^{12}$C cross sections for three additional kinematical setups, corresponding to incoming energies $E_e=620$ MeV, $E_e=961$ MeV, $E_e=1650$ MeV and scattering angles $\theta_e=60^\circ$, $\theta_e=37.5^\circ$, and $\theta_e=11.95^\circ$, respectively. In all cases, we observe an overall good agreement between data and our full theoretical model. Analyzing the separate contributions of the different elementary reaction mechanisms it clearly emerges that including the ANL-Osaka DCC pion-production amplitudes is crucial to reproduce experimental data in the resonance region. Consistently with Ref.~\cite{Rocco:2018mwt}, MEC are of primarily transverse nature and are needed to fill the missing strength between the $\Delta$ and the QE peaks. There are three main missing  ingredients in our framework that are responsible for the relatively small discrepancies with experimental data. In this work we have neglected the interference between one- and two-body currents, which has been proven to enhance the QE peak of the transverse response function~\cite{Benhar:2015ula,Lovato:2016gkq}. In addition, the treatment of FSI in two-nucleon emission processes is not as accurate as in the one-nucleon knockout case, whereas for real-pion production they are neglected altogether. Finally, at variance with the MEC, the ANL-Osaka DCC amplitudes do not encompass any in-medium modifications of the $\Delta(1232)$. More generally, it has to be noted that the MEC employed in this work were derived in Ref.~\cite{Simo:2016ikv} based on the weak pion-production model of Ref.~\cite{Hernandez:2007qq}. Despite the latter provides pion-production rates off the nucleon that are in good agreement with those of the ANL-Osaka DCC model~\cite{Sobczyk:2018ghy}, we are making efforts to employ MEC that are consistent with the ANL-Osaka DCC amplitudes.  

Within the same framework adopted to study inclusive electromagnetic scattering, we have carried out calculations of the double-differential CC $\nu_\mu$-$^{12}$C scattering cross sections for
$E_\nu=1$ GeV, $\theta_\mu=30^\circ$ and $E_\nu=1$ GeV, $\theta_\mu=70^\circ$. As expected, real-pion emission provides significant excess strength in the $\Delta$ peak, while MEC primarily contribute in the dip region. In view of the above-mentioned limitations, we refrain from computing the flux-folded differential cross sections, which could be readily compared to experimental data.  Work in this direction is underway, and, together with a further extension of the factorization scheme to account for two-pion emission processes will be the subject of future works.

\section{Acknowledgments}
This research is supported by the U.S. 
Department of Energy, Office of Science, Office of Nuclear Physics, under contracts DE-AC02-06CH11357,
by Fermi Research Alliance, LLC, under Contract No. DE-AC02-07CH11359 with the U.S. Department of Energy, Office of Science, Office of High Energy Physics,
and by National Natural Science Foundation of China (NSFC) under contracts 11625523.
Numerical calculations have been made possible through a CINECA-INFN  agreement, providing access to resources on MARCONI at CINECA.

\bibliography{biblio}

%%%% THE FOLLOWING MUST BE INCLUDED IN THE biblio.bib file

%\bibitem{tjon}
%J.Tjon

%\bibitem{Horikawa:1980cv}
%F.Lentz and M. Thies

%\bibitem{chen-lee}
%C. Chen and T.-S. H. Lee

%\bibitem{benhar}
%Benhar

%\bibitem{sato}
%sato

%\end{thebibliography}{}

\end{document}